\documentclass[twocolumn]{aastex6}
\usepackage{amsmath,graphics,epsfig,xcolor}
\usepackage{natbib,hyperref}
\usepackage[referable]{threeparttablex}

\def\teff   {{$T_{\rm eff}$}}

\def\msun   {{M$_{\odot}$}}
\def\rsun   {{R$_{\odot}$}}

\def\fspot  {{$f_{spot}$}}
\def\xspot  {{$x_{spot}$}}
\def\fblack {{$f_{black}$}}

\defcitealias{somers2015a}{Paper I}

\begin{document}

\title{The SPOTS Models: A Grid of Theoretical Stellar Evolution Tracks and Isochrones For Testing The Effects of Starspots on Structure and Colors}

\email{garrettsomers@gmail.com}

\author{Garrett Somers\altaffilmark{1,2},
        Lyra Cao\altaffilmark{3},
        Marc H. Pinsonneault\altaffilmark{3,4}}
\altaffiltext{1}{Department of Physics \& Astronomy, Vanderbilt University, 6301 Stevenson Center Ln., Nashville, TN 37235, USA}
\altaffiltext{2}{VIDA Postdoctoral Fellow}
\altaffiltext{3}{Department of Astronomy, The Ohio State University, Columbus, OH 43210, USA}
\altaffiltext{4}{Center for Cosmology and Astroparticle Physics, The Ohio State University, Columbus, OH 43210, USA}

\begin{abstract}

One-dimensional stellar evolution models have been successful at representing the structure and evolution of stars in diverse astrophysical contexts, but complications have been noted in the context of young, magnetically active stars, as well as close binary stars with significant tidal interactions. Numerous puzzles are associated with pre-main sequence and active main-sequence stars, relating to their radii, their colors, certain elemental abundances, and the coevality of young clusters, among others. A promising explanation for these puzzles is the distorting effects of magnetic activity and starspots on the structure of active stars. To assist the community in evaluating this hypothesis, we present the Stellar Parameters Of Tracks with Starspots (SPOTS) models, a grid of solar-metallicity stellar evolutionary tracks and isochrones which include a treatment of the structural effects of starspots. The models range from 0.1-1.3\msun and from spot-less to a surface covering fraction of 85\%, and are evolved from the pre-main sequence to the red giant branch (or 15~Gyr). We also produce two-temperature synthetic colors for our models using empirically-calibrated color tables. We describe the physical ingredients included in the SPOTS models and compare their predictions to other modern evolution codes. Finally, we apply these models to several open questions in the field of active stars, including the radii of young eclipsing binaries, the color scale of pre-main sequence stars, and the existence of sub-subgiants, demonstrating that our models can explain many peculiar features of active stars.

\end{abstract}

\section{Introduction}

Theoretical models of the structure and evolution of stars are enormously important for interpreting observable stellar properties. As such, a large number of suites of theoretical evolutionary tracks and isochrones have been produced over the years and are available in the literature. The vast majority of these calculations can be described as ``standard stellar models'', meaning that they neglect the impact of phenomena such as stellar rotation, magnetism, binary interactions, and mass loss.  This framework has proved quite successful in predicting the behavior of stars in a wide variety of astrophysical contexts. However, the success of standard models has not been universal. One era of stellar evolution where models have been less accurate is the pre-main sequence, where a number of physical effects which are less pronounced or absent in older systems influence the evolutionary trajectory of stars: accretion from circumstellar material, magnetic interactions with their proto-planetary discs, powerful dynamo-generated magnetic fields, strong starspot activity, and rapid rotation. Such effects cannot be addressed in the standard model framework.

Unsurprisingly then, a number of problems arise when comparing the predictions of theoretical models to the observed properties of young stars. To name a few: the inferred ages of pre-main sequence stars can differ significantly between objects within the same cluster \citep[e.g.][]{pecaut2012,malo2014,herczeg2015,pecaut2016,feiden2016}; the colors predicted by model atmospheres differ in peculiar ways from observations \citep{gullbring1998,pecaut2013}; and the observed lithium abundance patterns of young clusters disagree with the predictions of standard stellar models \citep{soderblom1993a,somers2014,somers2015b,jeffries2017} in a manner that cannot be reconciled without invoking the physical processes that standard models neglect \citep{pinsonneault1997}.  These modeling discrepancies are problematic. Numerous questions of astrophysical importance rely on the accurate modeling of young stars, such as the circumstellar disk and exoplanet evolution timescales, studies of the initial mass function (IMF) of star formation, and galactic chemical evolution.

Rotation and magnetism are clearly important for young stars, and their inclusion is, in our view, physically well-motivated. Rotation induces mixing, important for older stars, and changes stellar structure. However, because stars spin down as they age, the departure from spherical symmetry is modest, leading to relatively small direct structural effects. As a result, in the young star context the direct effect of rotation is real but expected to be modest.

Stellar activity induced by magnetism, by contrast, can have a substantial impact on both the interpretation of observables (such as colors) and theoretical models for cool stars. The surface of a spotted star does not have a unique temperature, and because the fraction of the surface area covered in spots can be large, these effects can be substantial. Even early on, the colors of active stars were seen to be different from those of inactive counterparts \citep{campbell1984}; these differences are large enough that distinct empirical color-mean surface temperature relations are needed for active and inactive cool stars \citep{pecaut2013}. Magnetic fields are known to inhibit convection in sunspots, and they could also clearly impact energy transport, and by extension stellar structure, in heavily spotted stars. A number of papers have incorporated models of magnetic effects into stellar interiors models to establish their predicted influences \citep[e.g][and references therein]{spruit1982,spruit1986,mullan2001,chabrier2007,feiden2013,jackson2014,somers2015a,azulay2017}. These papers have provided explanations for a number of outstanding problems with pre-main sequence stars.

In \citet[][Paper I hereafter]{somers2015a} we introduced our method for including starspots into evolutionary calculations and presented preliminary models. The key effects considered were the modification of surface boundary conditions from an inhomogeneous surface temperature, flux blocking by spots in convective regions, and the impact of large filling factors on the observed colors of stars. From our tests, we concluded that intensive magnetic activity on the pre-main sequence can explain a number of pre-main sequence properties that are challenging for standard models, including mass-dependent age gradients in young clusters, the anomalous colors of some rapidly rotating young stars, and spreads in the surface abundance of Li between stars of equal mass and age. In some cases, such as lithium depletion, we argued that there could be signals of \textit{prior} strong starspot effects even in stars with current modest spot filling factors. In this paper, we expand upon this previous work through updates to our spot methodology and predictions for the colors of spotted stars, and present the Stellar Parameters Of Tracks with Starspots (SPOTS) models, a series of evolutionary tracks and isochrones for the community\footnote{\url{https://zenodo.org/record/3593339}}. These models are timely, as there has been significant interest in empirical studies of starspots on young stars in recent years \citep[e.g.][]{fang2017,gully-santiago2017,guo2018,rackham2018,morris2018}. These models include the impact of starspots on the theoretical HR Diagram position, as well as on colors, and include a range of spot properties.

Our paper is arranged as follows. In $\S$2 we discuss our treatment of starspots, the adopted physical ingredients for our models, and our method for producing model photometry. In $\S$3, we describe the grid of models we have produced and compare them to literature models. In $\S$4, we discuss a number of potential applications of our spot models, including comparisons of our models with pre-main sequence eclipsing binaries, sub-subgiants, and the colors of young stars, as well as discussing the model-derived masses and ages of a young cluster. We summarize our findings and conclude in $\S$5.

\section{The Models}\label{sec:themodels}

This section discusses the physical ingredients used in our evolutionary calculations, starting with our implementation of starspots, followed by our chosen microphysics, how we produce model photometry for our tracks, and ending with a discussion of initial conditions. Briefly, our models account for starspot-induced flux-blocking in the interior of the star, and the influence of the altered surface temperature on the surface boundary conditions. These effects have structural consequences for our models, which we self-consistently account for. We also produce two-temperature colors from empirical tables, by considering the emission from the hot and cool regions of the stellar surface.

\subsection{Starspot treatment}\label{sec:spots}

The influences of starspots on the structure of our models is detailed in \citetalias{somers2015a}, and we refer the reader to Section 2.2 therein for detailed discussions. Briefly, the effects of starspots on our models are to both alter the average pressure and temperature at the model photosphere, and to suppress the rate of convective energy transport in the sub-surface layers. The spot properties in our models are characterized by a total spot filling factor (\fspot) and the ratio of the temperature of the spots to the temperature of the warm ambient regions (\xspot), which are free parameters. The temperature at a given layer in the star is determined by summing the fluxes of the hot and cool regions:

\begin{equation}\label{eqn:temps}
    T^4 = (1-f_{spot}) T_{hot}^4 + f_{spot} T_{cool}^4
\end{equation}

\noindent
where $T_{hot}$ and $T_{cool}$ are the local thermodynamic temperatures of the hot ambient and cool spot regions, respectively. When evaluated at the photosphere, Eq. \ref{eqn:temps} defines the effective temperature of the star. \fspot\ and \xspot\ can take values between 0 and 1, but within our code any combination of the two parameters corresponds to an equivalent filling factor of pure black spots (\fblack), defined by

\begin{equation}\label{eqn:fblack}
1 - f_{black} = (1 - f_{spot} + f_{spot} x_{spot}^4)
\end{equation}

Eq. \ref{eqn:fblack} implies that a larger starspot fraction of warmer spots is equivalent to a lower starspot fraction of cooler spots from the point of view of the structural variables. However different spot fractions will have different implications for synthetic colors and magnitudes (see $\S$\ref{sec:colors}). We calculate equivalent models with filling factors of pure black spots (\fblack) of 0.0-0.5 in steps of 0.1. Following \citetalias{somers2015a} we adopt \xspot\ = 0.8, which is characteristic of the observed range of starspot temperature ratios in young stars and sub-giants \citep{berdyugina2005}. With this value of \xspot, our chosen series of \fblack\ models equates to \fspot\ values of of 0.00, 0.17, 0.34, 0.51, 0.68, and 0.85.

In the interior, spots inhibit the local flux near the surface layers, rerouting radiative transport to the non-spot regions by a factor proportional to the spot intensity. We treat this by enhancing the radiative flux in the convective layers by the factor given in Eq. \ref{eqn:rad}:

\begin{equation}\label{eqn:rad}
    \nabla_{\rm rad,spot} = \nabla_{\rm rad} / (1 - f_{black})
\end{equation}

Although the fields we model are quite large, they are not susceptible to the magnetic buoyancy which can dissipate deep internal magnetic fields \citep[e.g][]{browning2008,yadav2015}, as the primary structural influence occurs at very shallow depths where buoyancy is inefficient. 

For the purposes of setting the outer boundary condition at the photosphere, we assume spot and non-spot regions are in pressure equilibrium. The consequence is that the pressure of the hot surface regions (solely gas pressure) is equal to the pressure in the spot regions (the sum of gas and magnetic pressures). To define the relation between surface pressure, effective temperature, and surface gravity, we use the model atmospheres discussed in $\S$\ref{sec:microphysics}. These model atmospheres are in effect look-up tables where you input a surface gravity and an effective temperature, and get back a photospheric pressure. We use the pressure in the hot (non-magnetic) regions and the gravity at the surface for this purpose. This table look-up is summarized in Eq. \ref{eqn:surf}:

\begin{equation}\label{eqn:surf}
    P_{\rm surf} = P_{\rm surf}(T_{\rm hot, surf},\log g)
\end{equation}

\noindent
where $P_{\rm surf}$ is the surface pressure, $T_{\rm hot, surf}$ is the temperature of the non-spot photosphere regions, and $\log g$ is the surface gravity.

Ours is not the only formulation for testing the influence of magnetic activity on stellar structure, and we briefly mention some predecessors (for more discussion, see \citet{somers2015a}). Our treatment bears the most similarity to the starspot models of \citet{spruit1986}, whose work inspired our efforts. However, we expand on their work by allowing varying starspot temperatures instead of modeling pure black spots, and we also model different evolutionary states. \citet{chabrier2007} present calculations of the influence of a reduced mixing length and cooled stellar surface arising from pure black starspots, and achieve results qualitatively similar to ours. However, our use of variable temperature spots permits more realistic color predictions. In a series of papers, \citet{mullan2001} and collaborators have explored the consequences of strong internal magnetic fields and/or spots on structure. Finally, \citet{feiden2013} have produced fully-consistent stellar evolutionary calculations employing a strong internal magnetic field. We compare our results to these in $\S$\ref{sec:grid}. Each of these works have produced important results, and we consider our efforts a continuation of this line of study.

\subsection{Microphysics}\label{sec:microphysics}

Our input physics is similar to that used in our previous published work, with changes primarily in the treatment of surface boundary conditions and the transformation from the theoretical to the observational plane.  We begin with the physics in common and then follow with the newer ingredients. We adopt nuclear reaction cross-sections from \citet{adelberger2011}, and deuterium burning is included in our models.  Our high temperature opacities are from the OP project \citet{mendoza2007}, using the \citet{grevesse1998} mixture of heavy elements favored by helioseismic data and a surface Z/X value of 0.0231. Low temperature molecular opacties are from \citet{ferguson2005}. We use the OPAL equation of state \citep{rogers1996,rogers2002}. Although our grid extends to higher mass stars, microscopic diffusion is not included in these calculations. The interplay between diffusion, radiative levitation, and rotational mixing is quite complex and outside the scope of our modeling efforts. This exclusion will not impact our results because the physical effects of gravitational settling are quite small in young stars, which are the primary focus of this work.

We calibrate the mixing length and the initial hydrogen, helium and metal content (X,Y,Z) adopting a solar radius, luminosity and age of $6.9598 \times 10^{10}$~cm, $3.8418 \times 10^{33}$~erg s$^{-1}$, and 4.568~Gyr respectively.  Our final solar values are $\alpha = 1.807$ and (X,Y,Z) = (0.7195,0.2676,0.0165).

We employ two different sets of model atmospheres for the higher and lower mass ranges of our grid. Below 0.4\msun\ we use the Allard \citep{allard1997} model atmospheres, and above 0.6\msun\ we use the Castelli-Kurucz model atmospheres \citep{castelli2004}. This division is born of necessity -- the Allard models do not extend to low surface gravities, thus preventing sufficient post-main sequence calculations, and the Castelli models do not consider effective temperatures below 3500K, thus preventing calculations at the low-mass end. We have elected to use these grids as they were the dominant atmospheres available at the time the work was done. In particular, for purposes like computing the convective overturn timescale, models which adopt the \citet{grevesse1998} mixture are closer to reproducing the solar convection-zone properties, so in the absence of new interiors physics (such as opacities), it’s reasonable for our purposes to adopt that mixture.

At fixed mass, the two model atmospheres give temperatures differing by 50-100~K at fixed radius, with larger discrepancies on the pre-main sequence than the main sequence. Consequently, an abrupt change from one model atmosphere to the other would cause a discontinuity in our isochrones and tracks. To address this concern, we construct a ramp between 0.4 and 0.6\msun\ from the Allard models to the Castelli models. We set 0.35\msun\ and 0.65\msun\ as the bounds of the ramp, the former using solely the Allard model and the latter using solely the Castelli model. For each intervening mass step, we calculate models using both atmospheres and linearly interpolate at fixed age between the two to produce an intermediate model. The distance interpolated between the two models for each mass point corresponds to where the mass point is relative to the bounds of the ramp -- for example, 0.4\msun\ interpolates 1/6 of the distance between Allard and Castelli, 0.45\msun\ interpolates 2/6 of the distance between Allard and Castelli, and so forth. In this fashion, the transition from the lower mass end to the higher mass end is smoothed out.

\subsection{Color Transformations}\label{sec:colors}

In additional to structural variables, photometric magnitudes and colors are a useful comparison point between published tracks and stellar observations. We have chosen to adopt empirical relations for converting the $L$ and \teff\ of our models to colors. While this restricts the range of calibrated parameter space relative to analytic colors from model atmospheres, we find that empirical relations produce colors which closer reflect nature and are thus more likely to be useful to observers. 
For this purpose, we adopt the empirical main sequence color transformations, originally published in \citet{pecaut2013} and continuously updated online\footnote{\url{http://www.pas.rochester.edu/~emamajek/EEM_dwarf_UBVIJHK_colors_Teff.txt}}. From their tables, we take the V-band bolometric corrections as a function of luminosity, and the \teff-color relations for $B$, $V$, $R_C$, $I_C$, $J$, $H$, $K_S$, $W1$, $W2$, and the new {\it Gaia} DR2 bands $G$, $BP$, and $RP$ \citep{gaiadr1_2016,gaiadr2_2018}.

We next calculated colors and magnitudes in all of the above bands for our un-spotted tracks and isochrones. To do this, we convert the model luminosity directly into a V-band magnitude using the appropriate bolometric corrections. We then convert the model \teff\ into every color which has a V-band value in it (i.e. $B-V$, $V-Rc$, $V-G$, etc.). The absolute magnitude in each band in then the sum (or difference) of the color and the absolute V-band magnitude. 

For the spotted stars, the process is more complicated as we wish to produce colors and magnitudes which are the sum of the emission from the hot and cool regions. To do this, we first collect the radius and the hot and cool surface temperatures for each model step. From these values, we determine the colors and magnitudes that would emanate from stars of that radius and with those respective surface temperatures. Finally, we add together the fluxes from the two mock stars, weighting by the flux difference and by the relative surface area of the hot and cool regions. Note that these colors assume that the observed light is a sum of the hot and cool regions, in direct proportion to their total covering fraction. In real spotted stars, asymmetry implies that some hemispheres may have different covering fractions that the total surface covering fraction, so the published colors should be thought of as averages over the full surface.

\begin{figure*}
\centering
\includegraphics[width=0.9\linewidth]{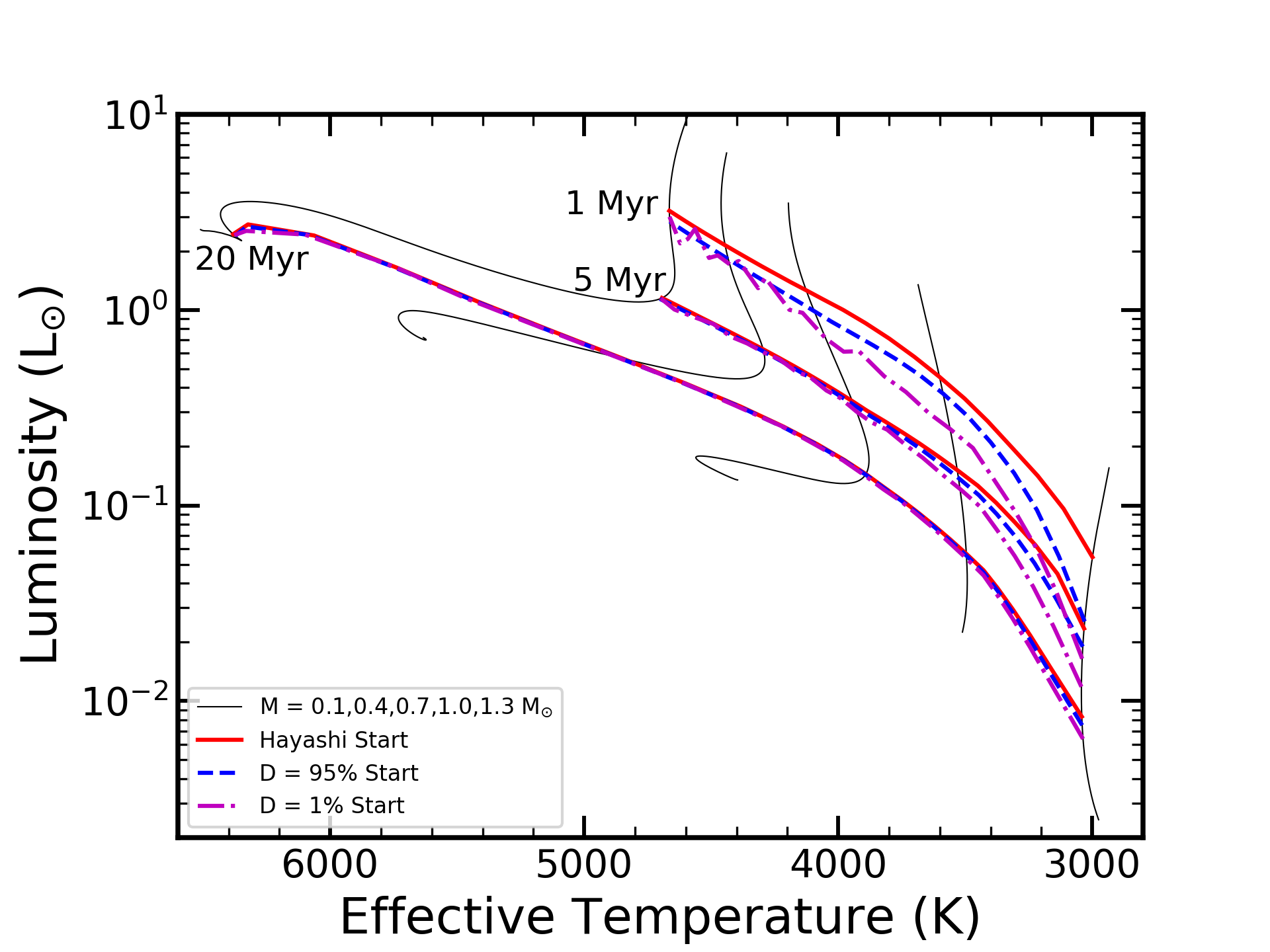}
\caption{The influence of initial conditions on the location of isochrones for young stars. Red isochrones begin the clock high above the Hayashi track. Blue isochrones begin the clock at the moment when 95\% of initial deuterium remains. Purple isochrones begin the clock at the moment when 1\% of initial deuterium remains. These different choices affect the location of stellar models at early ages on the low-mass end.
\label{fig:DBL}
}
\end{figure*}

\subsection{Model Initial Conditions}


The setting of initial conditions for pre-main sequence stellar models is an old and tricky problem for a few reasons. First and foremost, star-formation is a fundamentally continuous process passing through different stages from initial molecular cloud collapse to the onset of core hydrogen burning. Which stage in this process one calls the beginning of stellar evolution is not obvious or well-defined. Second, one-dimensional models generally assume that the totality of the stellar mass has been accreted at the start, when in fact mass is built up by an initial stellar core through repeated episodes of mass ingestion from the surrounding circumstellar material \citep{vorobyov2005,evans2009,vorobyov2015}. Consequently, the evolution of very young stars through the H-R diagram is not simply a clean descent along the Hayashi track, but also includes a stochastic component related to accretion bursts, and a long-term trend towards hotter and more luminous as the total mass increases \citep[e.g.][]{baraffe2009}. Fundamentally, proto-stars appear to traverse unique paths in the young H-R Diagram due to these processes -- there is support for this in the luminosity spreads observed in young associations \citep[e.g.][]{hillenbrand1997,hartmann2001,dario2010}. Our models do not treat this accretion phase, and begin high up on the Hayashi track with their full mass accumulated.

Historically, modelers have developed a few conventions for t=0, which have their own advantages and disadvantages. One approach is to initialize a stellar model that is very far up the Hayashi track and define its starting point as the birth time of the star. These models have very high luminosity and a large radius, making their instantaneous Kelvin-Helmholtz timescale quite short. Consequently, they have a brief initial phase of rapid contraction, and subsequently settle into a slower pre-main-sequence evolution. This has two advantages: 1) simplicity; 2) insensitivity to the precise physical choices for the initial model. This second point is because the initial contraction phase is so short (t $\lesssim$ 10$^5$ yrs) that a poor choice of initial model will converge to a physically reasonable one quickly, and the elapsed time is negligible even on pre-MS timescales. However, this choice does not peg the initial time to a true physical stage in proto-star evolution, and the initial portion of the track is clearly unphysical.

Another common approach is to consider the burning of deuterium, which stars are born with and which is rapidly astrated on the early pre-MS \citep[e.g.][]{stahler1988}. Modelers may elect to define the start of the pre-main sequence as the onset of deuterium burning or the completion of deuterium burning, and there are trade-offs with these two choices. The advantage of beginning the clock at the start of the so-called deuterium birth-line (DBL) is that this is a physically-motivated choice. As stars continue to accrete matter, the newly-engulfed deuterium is burned in the stellar center, thus injecting new energy into the stellar interior. This extra heat counteracts normal Kelvin-Helmholtz contraction, and temporarily stalls the standard pre-MS collapse at the top of the DBL. Once the heavy accretion phase subsides, stars will destroy their remaining deuterium and progress down the pre-MS. This epoch thus provides a logical beginning point to normal Hayashi contraction. However, the rate of deuterium burning is mass dependent --- higher-mass stars deplete their deuterium more rapidly ($\sim 10^5$ yrs) than low mass stars ($\sim 10^6$ yrs) once accretion ceases. Therefore this choice of initial conditions produces significant mass-dependent trends in the onset of Hayashi contraction within individual stellar populations -- i.e. high mass stars begin their standard evolution earlier than low-mass stars. This is disadvantageous for producing isochrones and studying stellar associations. The actual early history could also depend on both the mass dependence of the mean accretion rate on the birthline (which could allow massive stars to form first, or low mass ones, of even have all form simultaneously); this phenomenon is outside of the scope of this paper, as it involves processes in the hydrodynamic collapse phase.

To alleviate this complication, one may also choose to define the end of the DBL -- i.e. the time at which the last of the deuterium is destroyed -- as the beginning of the pre-main sequence. While this choice overcomes the non-coevality issue discussed above, it is arguably less accurate for low-mass stars which take significant time to destroy their deuterium.

To illustrate the impact of these different choices of initial conditions, we show isochrones produced with each of the three choices discussed above in Fig. \ref{fig:DBL}. In this figure, the thin grey lines are evolutionary tracks for several different masses. The red, blue, and magenta lines are 1, 5, and 20 Myr isochrones constructed by initializing t=0 at, respectively, a location far above the Hayashi track, the time when 95\% of initial deuterium remains, and the time when 1\% of initial deuterium remains. 

For 1~Myr and 5~Myr, the isochrones are quite similar at the high-mass end, but diverge substantially in the M-dwarf regime. Notably, the 5~Myr 0.1\msun\ stars initialize at the end of the DBL actually appear younger than the 1~Myr 0.1\msun\ stars initialized at the beginning of the DBL. This demonstrates that these initial choices have significant consequences for interpreting stars at the low mass end. However, by 20~Myr these discrepancies have nearly vanished because of the increasingly slow rate of pre-MS contraction. Note that in each case the model runs are identical, it is merely the time at which t=0 has been defined that varies.

For this paper, we have chosen the first of these three options, and initialize our models far up the Hayashi track. Because we understand that others may prefer a different initialization strategy, we have included in our tracks deuterium abundance as a parameter. The ages of the models can be easily re-scaled to select various milestones along the DBL.

\begin{figure*}
\centering
\includegraphics[width=0.9\linewidth]{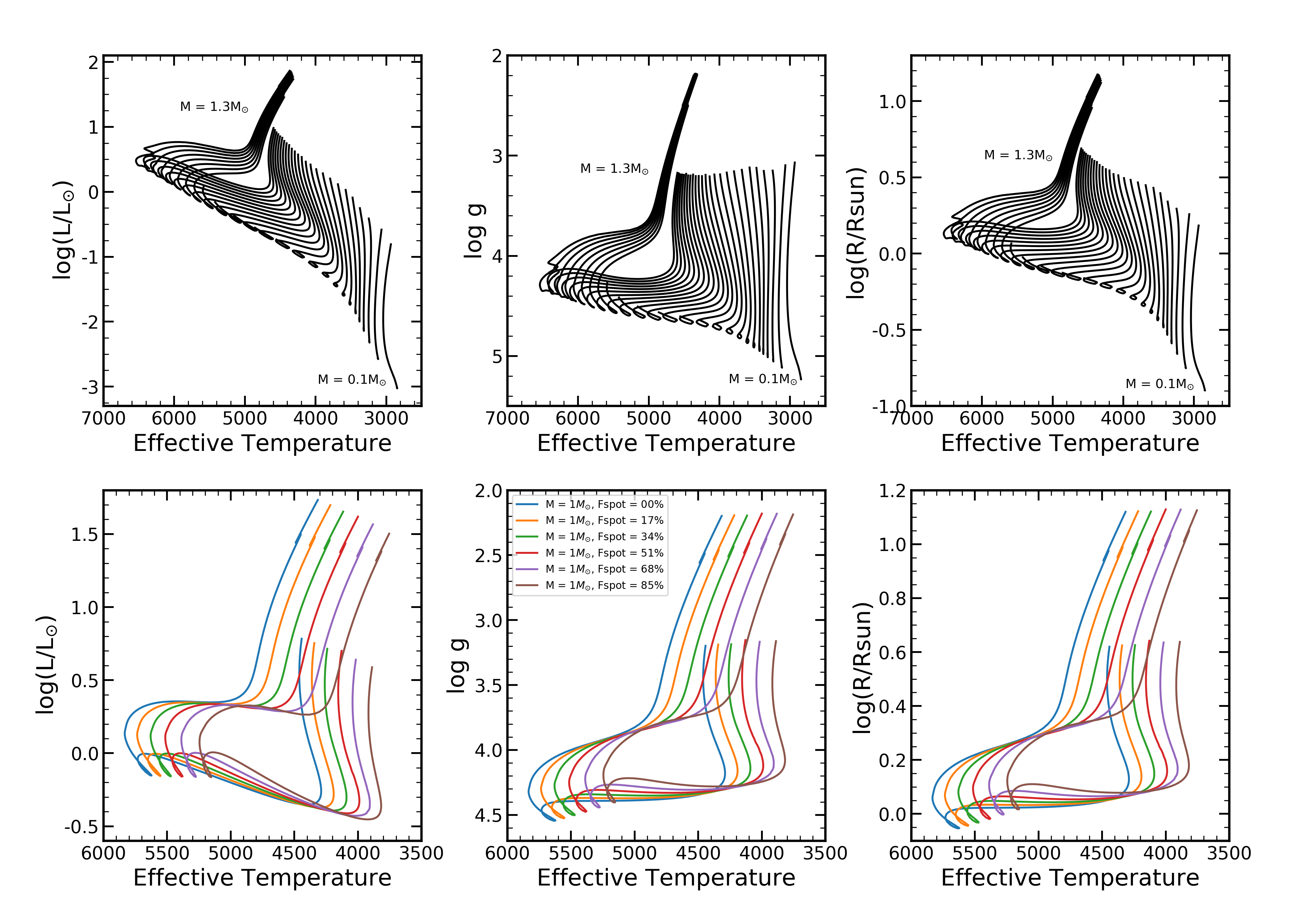}
\caption{Individual stellar tracks from the SPOTS models. {\it Top row:} Diagrams of the luminosity (left), surface gravity (center), and radius (right) versus \teff\ for our un-spotted models from 0.1-1.3\msun. {\it Bottom row:} An illustration of the influence of starspots on the a 1\msun\ stellar model. From blue to brown, the starspot intensity increases, as indicated by the caption in the center panel. The primary effect of spots is to reduce \teff\ and increase the radius by up to $\sim 15$\%.
\label{fig:StellarTracks}
}
\end{figure*}

\begin{figure*}
\centering
\includegraphics[width=0.9\linewidth]{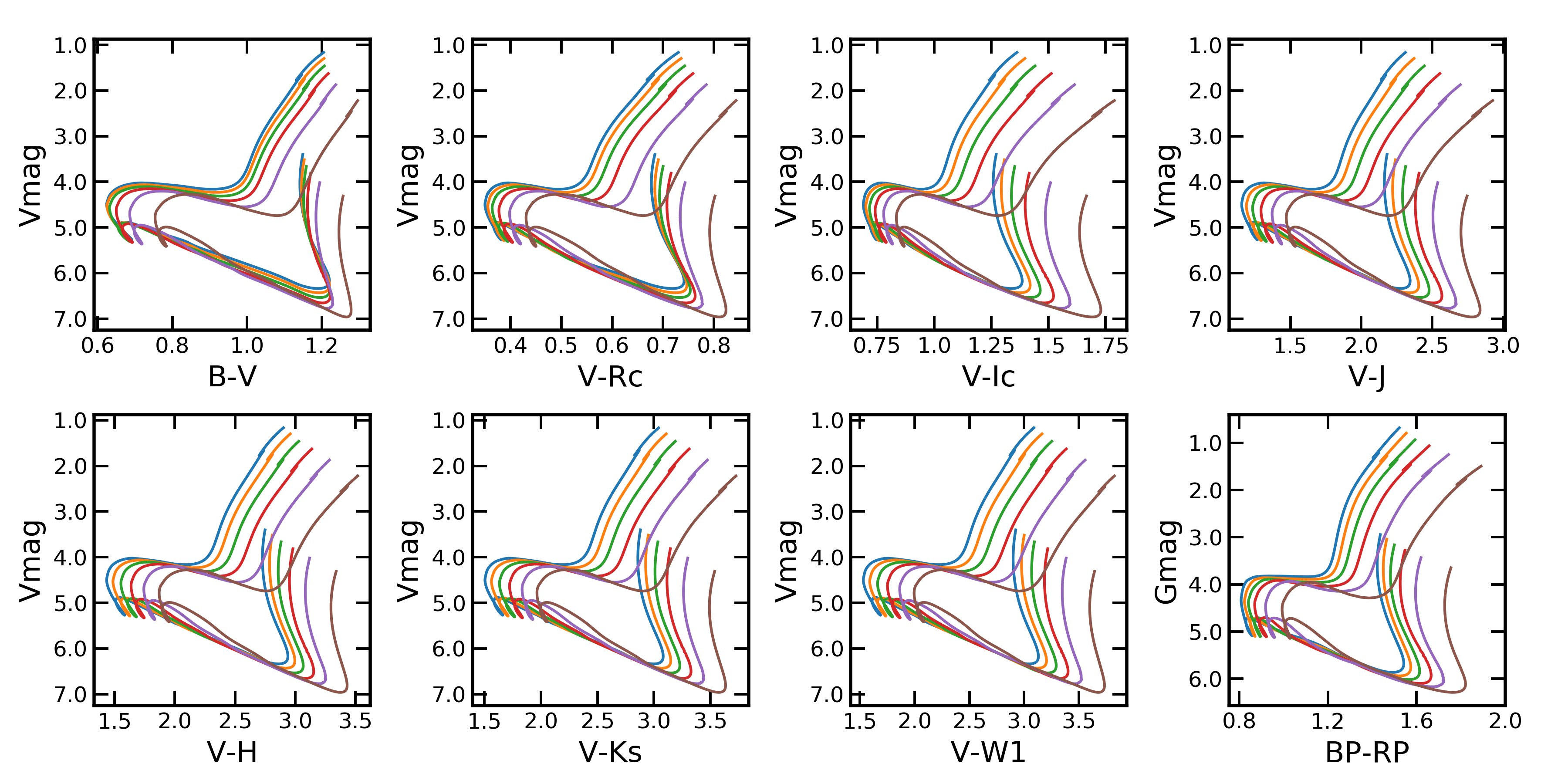}
\caption{The influence of starspots in several different color bands from the SPOTS models. Each track is 1\msun\ and the different colors reflect the different spot intensities given in the caption of Fig. \ref{fig:StellarTracks}. The influence is different in each band, reflecting the fact that different colors sample different portions of the changing S.E.D. of spotted stars.
\label{fig:ColorTracks}
}
\end{figure*}

\begin{figure*}
\centering
\includegraphics[width=0.9\linewidth]{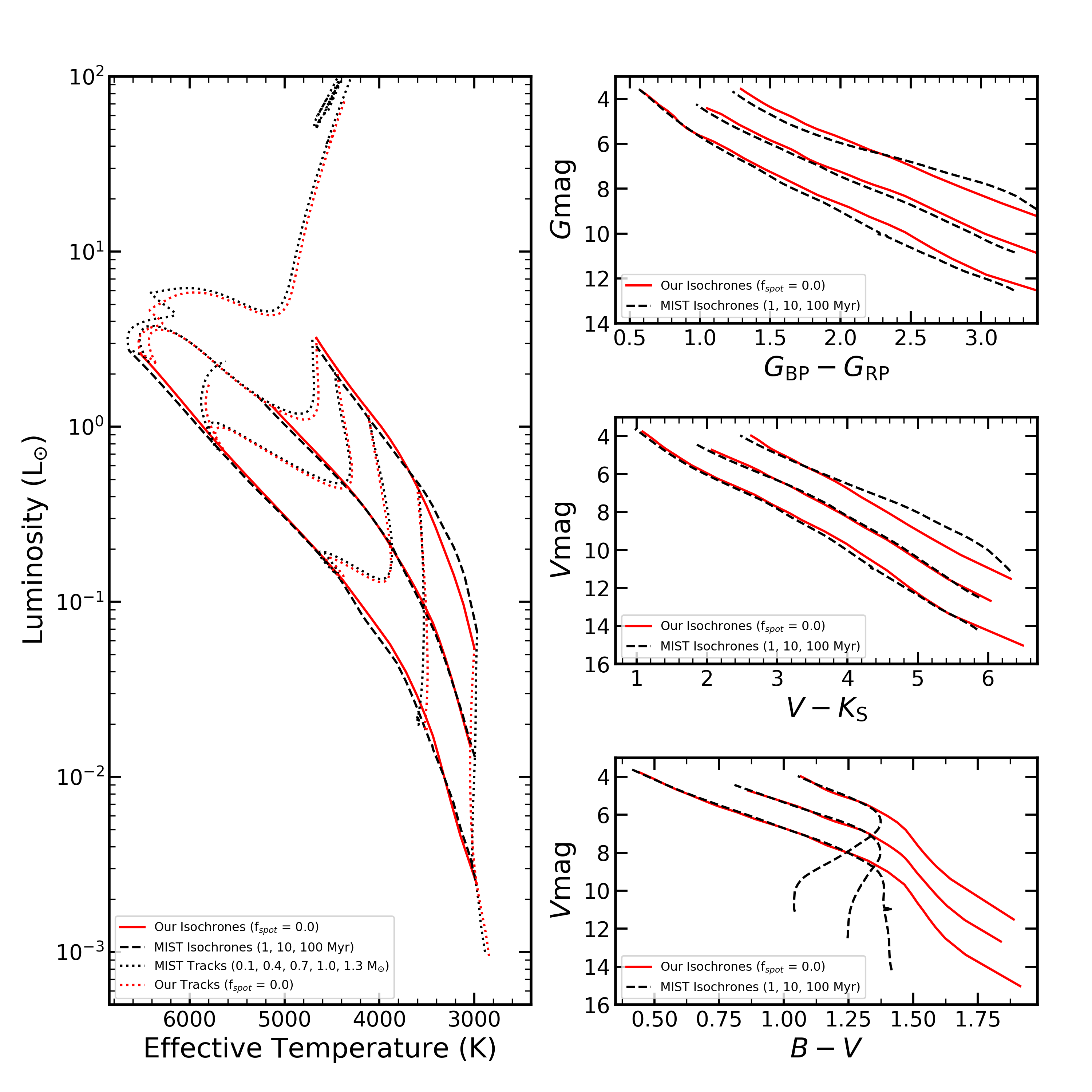}
\caption{A comparison of the non-spot SPOTS models to those of the MIST collaboration \citep{dotter2016,choi2016}. {\it Left:} Three isochrones and four evolutionary tracks from SPOTS (red) and MIST (black). Overall the agreement is quite good at early times. {\it Top Right:} A comparison of the three isochrones from the left panel in the {\it Gaia} CMD. The agreement is good at higher masses, but we find substantial departures towards the low mass end driven by the different calibrations. {\it Middle Right:} Same as top right, except for $V$ vs $V-K_S$. The agreement is to that seen in the {\it Gaia} CMD. {\it Bottom Right:} Same as top right, but in $V$ vs. $B-V$. The agreement is excellent at higher temperatures, but we find very large departures below $B-V = 1.3$. This results from complications in the MIST color calibrations at the blue end of the spectral energy distribution for low mass stars.
\label{fig:MistComp}
}
\end{figure*}

\section{The Grid}\label{sec:grid}

Using the physical ingredients discussed in $\S$\ref{sec:themodels}, we compute tracks in the mass range from 0.1--1.3\msun. Every star is evolved until it surpasses the luminosity bump on the lower red giant branch, or until it reaches 15~Gyr, whichever comes first. We adopt a fixed spot-to-temperature ratio of 0.8, and calculate grids with spot covering fractions of 0.00, 0.17, 0.34, 0.51, 0.68, and 0.85 (see $\S$\ref{sec:spots}). From these models, we have produced evolutionary tracks at mass intervals of 0.05, and isochrones at a large number of logarithmically-spaced ages.

Examples of the evolutionary models in our grid appear in Fig. \ref{fig:StellarTracks}. First, we show in the top row only our non-spot models over the full mass range. Three different variables, luminosity, surface gravity, and radius, are compared to the \teff\ values of the models. As can be seen, models of 0.95\msun\ and above reach the post-MS by 15~Gyr. For these runs, we do not provide values once log~g of 2.1 has been exceeded.

The bottom row shows each 1.0\msun\ model in our grid, now varying the starspot covering fraction from 0--85\%. The impact of spots on our models is clear -- the tracks are shifted towards lower temperature, higher radius, and moderately lower surface gravity. The influence of activity on luminosity during the main sequence is negligible, though it becomes more pronounced on the red giant branch. These results confirm the predictions of previous magnetic starspot models, which have consistently found that heavily spotted stars have lower effective temperature and a larger radius of order 5-10\% \citet[e.g.][]{spruit1986,chabrier2007,jackson2014}.

An example of the colors we have calculated for these models are shown in Fig. \ref{fig:ColorTracks}. Here, we show B, Rc, Ic, J, H, Ks, and W1 relative to V in the first seven panels, and the {\it Gaia} CMD in the eighth. Once again the influence of starspots is clear, though the severity of the shifts show interesting color dependencies. For example, the main sequence locus of the first four tracks (\fspot\ = 0.00--0.51) lie one top of one another in B-V vs. V, but are more evenly spaced out in V-Ks vs. V. This results from the differences in the distortion of the spectrum sampled by the different bandpasses. These varying color shifts provide direct observational tests of the quality of our colors, which we consider in the following section.

Historically models of young stars have shown significant discrepancies between one another \citep[e.g.][]{stauffer2014}, but grids published in the last few years have converged significantly in their predictions \citep{herczeg2015}. An important test of the reliability of new stellar models is therefore a comparison with one of the newer established calculations of evolutionary tracks and isochrones. We present here a brief comparison between the zero starspot models from this paper and tracks and isochrones from the MIST collaboration \citep{dotter2016,choi2016}.

\begin{figure*}
\centering
\includegraphics[width=0.9\linewidth]{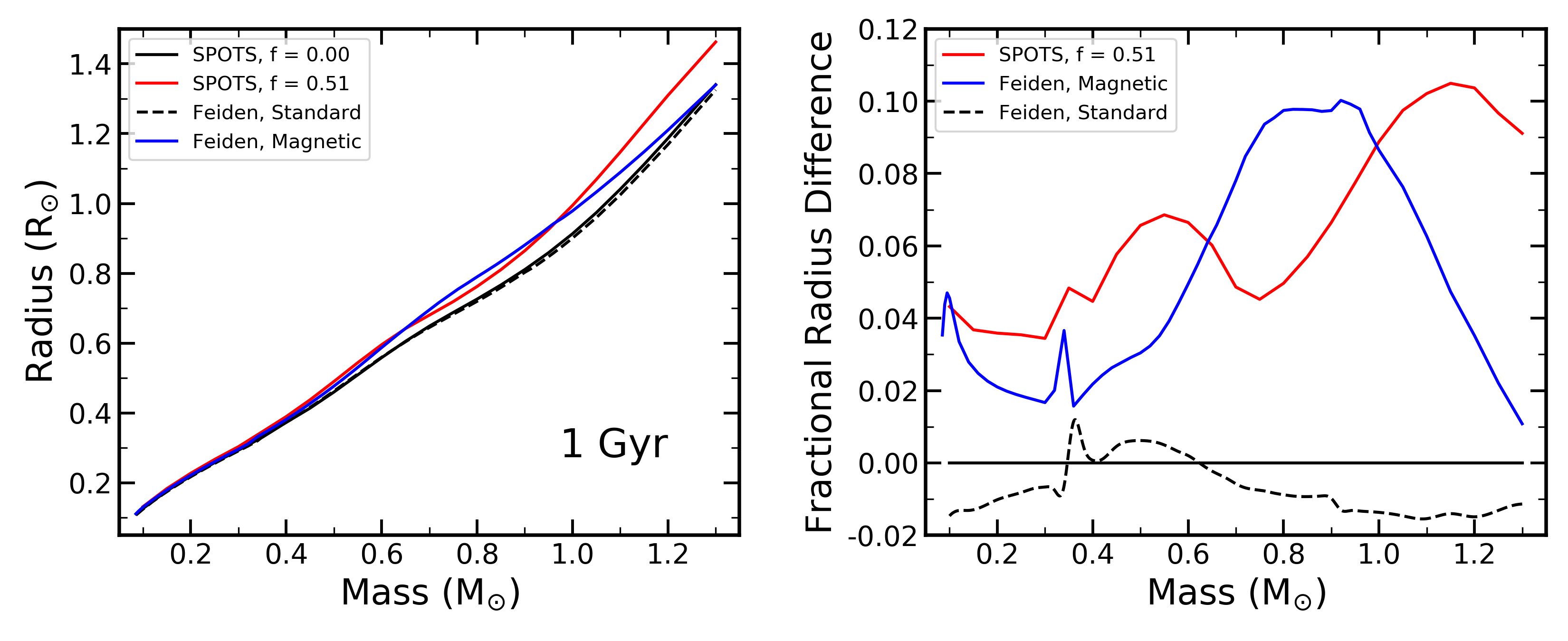}
\caption{A comparison between the SPOTS models and Feiden models at 1~Gyr. {\it Left:} The mass-radius relation of non-magnetic and magnetic versions of both SPOTS and Feiden models. {\it Right:} Same as the left, but detrended against the \fspot\ = 0.00 SPOTS model. Both magnetic models predict inflation rates of 3-10\%, but the morphology of the mass-dependent inflation percentage differs. Their similarity is encouraging given the many different choices made in constructing the two isochrones.
\label{fig:Feiden}
}
\end{figure*}

This comparison appears in Fig \ref{fig:MistComp}. First, on the left we show the theoretical H-R diagrams of our (red) and the MIST (black) models, with isochrones at 1, 10, and 100 Myr, and tracks of 0.1, 0.5, 0.9 and 1.3\msun. Overall the agreement is quite good, with the largest discrepancies appearing at the youngest age, as is typical. Slight post-MS discrepancies are likely due to the different abundance scales adopted between the two model sets -- \citet{grevesse1998} in the SPOTS models, and \citet{asplund2005} in the MIST models.

On the right, we compare the colors from our models to colors given in the MIST isochrones, again at 1, 10, and 100 Myr. Significant differences are evident in each panel, stemming mainly from the different methods of color calculation: we have interpolated from empirically-calibrated \teff-color relationships, whereas the MIST group generated model atmospheres for their tracks and integrated over the flux in the various bandpasses. In some cases the answers are quite similar -- the isochrones in $V$ vs. $V-K_{\rm s}$ lie nearly on top of one another at 10 and 100 Myr -- and in others the disagreement is worse -- the {\it Gaia} colors produce significant offsets at the cooler end. For $B-V$, the model atmospheres used by MIST to produce bolometric corrections (ATLAS12 and SYNTHE, see \citealt{choi2016} for details) have known issues reproducing the flux at the blue end, leading to significant disagreement cooler than $B-V \sim 1.3$. In this case the empirical color calibrations we have used should be more reliable.

A second interesting comparison point for the SPOTS models are independent grids of  stellar evolution calculations including the effects of magnetic fields. In a series of papers, \citet{feiden2013,feiden2014} introduced a suite of magnetically-active stellar models adapted from the Dartmouth evolution code. Their approach to modeling magnetic activity differs from ours in a few respects. As described in \citet{feiden2013}, the magnetic Dartmouth models incorporate Maxwell's Equations directly into the equations of stellar structure and define a radial magnetic field morphology to fill the B term in the equations. Moreover, their newest calculations (Feiden, personal communication; Feiden models hereafter) hold the surface magnetic field strength fixed at equipartition -- consequently, it changes over time with the evolving surface conditions. These techniques differ from ours both in model implementation and in how the magnetic field strength is defined and evolved. Nonetheless, a comparison between the two suites of models is instructive as to the range of theoretical expectations.

Fig. 5 presents this comparison. On the left, we show the mass-radius relationship of several models at 1 Gyr. First, the black solid and dashed lines are a non-magnetic isochrone for the SPOTS and Feiden models, respectively -- their close agreement suggests similar calibrations for the two suites, as is expected. The red and blue lines show, respectively, the 51\% spot models from the SPOTS grid and the magnetic models from Feiden. The right panel shows the same model, detrended against the non-magnetic SPOTS model. Here we see some important differences. At masses lower than $\sim 0.6$\msun, agreement is good between the two suites, predicting inflation percentages of $\sim$2-6\%. However, at higher masses, there are significant morphological differences that result from the disparate modeling approaches. The Feiden models reach a peak inflation percentage around 0.8\msun\ before declining almost to 0\% by 1.3\msun, while the SPOTS models continue to show higher inflation percentages up towards the high mass end. We note that at present there is no observational evidence for high spot filling factors in F stars, so this disagreement is in a regime of model space that is more theoretical than real. It is worth re-emphasizing that this comparison is not apples-to-apples, given the different prescriptions for the evolving surface magnetic field strength. However it is heartening that the results are reasonably consistent for lower mass stars. Moreover, the clear divergence in predictions above 1\msun\ may offer an interesting avenue for testing spot inflation vs. internal magnetic field inflation. 

\begin{figure*}
\centering
\includegraphics[width=0.9\linewidth]{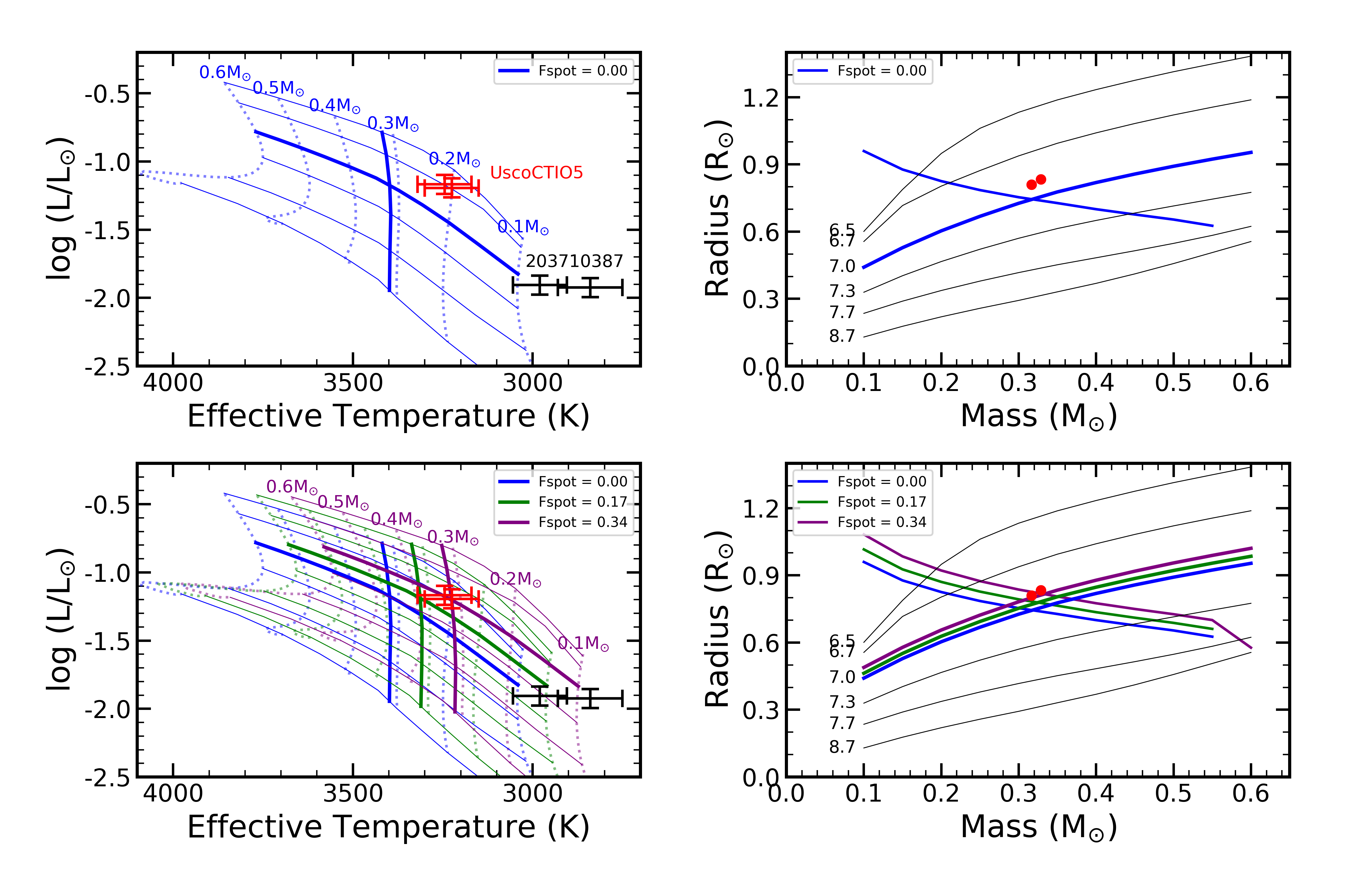}
\caption{Analysis of the eclipsing binary UScoCITO 5 and EPIC 203710387 \citep{kraus2015} with the SPOTS models. {\it Top Left:} The location of the four members of the two eclipsing binaries in the H-R diagram. The vertical blue line is a 0.32\msun\ spot-less model, equivalent to the mass of the binary components, and the horizontal solid line is an isochrone at 10~Myr, the assumed age of the Upper Sco association. {\it Top Right:} The location of USco5 in the mass-radius plane. The positive slope line is the expected mass-radius relation at 10~Myr, and the negative slope line is the locus predicted by the luminosity of the stars. In both top panels, it is clear that the models do not accurately predict the properties of USco5. {\it Bottom Left:} Same as top left, except models with \fspot\ = 0.17 and 0.34 have been added. The starspots cool the models and reduce the rate of Hayashi contraction such that the \fspot\ = 0.34 models correctly predict the \teff\ of the components of UscoCTIO5, and \fspot\ values of 0.2-0.4 predict the \teff\ of EPIC 203710387 at 10~Myr. {\it Bottom Right:} Same as top right, but models with \fspot\ = 0.17 and 0.34 have been added. The inflated radius induced by magnetic activity causes the mass-radius relation to move upwards in this plane, and the locus of constant-luminosity to also move upwards. The \fspot\ = 0.34 models now accurately predict the properties of USco5 in this plane.
\label{fig:KrausEB}
}
\end{figure*}

\section{Applications of SPOTS models}

\subsection{Radius inflation in pre-main sequence eclipsing binaries}

One of the primary motivations for constructing magnetic models is the widespread discrepancy between the radii of some young, active stars in eclipsing binary systems and model predictions. This discrepancy has been speculated as a consequence of incomplete atmosphere models, systematic errors in the radius measurements from eclipsing binaries, or a real effect arising from the interaction of magnetic activity and starspots with the physics governing the fundamental structure of stars \citep[][and references therein]{torres2010,feiden2012,stassun2014a,david2019}. A holistic comparison between the measured radii of stars and our models is beyond the scope of this paper, but we show here an demonstrative example of why starspot inflation is a promising solution to this problem.

\citet{kraus2015} studied a newly-discovered eclipsing binary UScoCITO 5 in the nearby Upper Scorpius association, first identified in data from the {\it Kepler} satellite's {\it K2} mission. By solving for the orbit, they precisely constrained the masses and radii of the two binary components, as well as their surface temperatures and luminosities. Through comparisons with published evolutionary models, the authors concluded that the measured parameters could not be reproduced at the assumed age of the association.

We have reproduced this comparison with our un-spotted evolutionary models in the top row of Fig. \ref{fig:KrausEB}, in the theoretical Hertzsprung-Russell diagram on the left and in the mass-radius plane on the right. The vertical lines on the left are evolutionary tracks at the specified masses, and the horizontal lines are isochrones at various ages from $\sim 3-500$~Myr. The thick blue lines are a single evolutionary track at the mass of the two components (model mass = 0.32\msun, components are 0.329\msun\ and 0.317\msun) and a single isochrone at 10~Myr, the putative age of the association \citep{feiden2016}. As can be seen, these models do not reproduce the properties of the eclipsing binary, which appears approximately 200~K cooler and 30\% more luminous than the models anticipate. Moreover, this discrepancy cannot be a product of an incorrect age of the association as the theoretical evolutionary track, fixed at the measured masses of the eclipsing binary components, remains too hot at any plausible cluster age.

On the upper right, we compare the mass and radius of the eclipsing binary components to a series of isochrones. The age of each isochrone is denoted by the number at its left end, in units of log(Age/yr). The positive-slope line is an isochrone at 10~Myr, and the negative-slope line shows the radius predicted by the models at each mass for a star whose luminosity at 10~Myr is equal to the measured value of the UScoCITO 5 components. Again we see that the data are larger than model predictions by $\sim$ 0.07\rsun or approximately 10\%. The discrepancies in both upper panels are consistent with the conclusions of \citet{kraus2015}.

While normal stellar models fail to predict the EB properties, the bottom row shows that our magnetic SPOTS models fare far better. In green and purple, we overlay tracks and isochrones at the same masses and ages for our \fspot\ = 0.17 and 0.34 models. In the bottom left panel, we see that the magnetic activity cools the stars at approximately fixed luminosity. When the models are cool enough to agree with the eclipsing binary components, the 10 Myr isochrone also remarkably runs right through the stars. Moreover, in the right panel, the inflated radius and moderately reduced rate of contraction for the \fspot\ = 0.34 starspot models results in a larger predicted radius at 10 Myr, once again agreeing remarkably well with the true values. We this see that the global predictions of inflated models are that pre-main sequence stars are cooler, older, and larger than normal stellar model predictions.

Our best-fit spot covering fraction of 0.34 agrees remarkably well with analysis performed by \citet{macdonald2017}, who fit the same eclipsing binary with their starspot models, finding a best-fit surface covering fraction of 0.33 $\pm$ 0.06. These authors performed a similar analysis on the eclipsing binary EPIC 203710387, another member of the Upper Scorpius association, using values derived by \citet{david2016}. The components of this system are lower mass ($M \sim$ 0.118 and 0.108 \msun) and smaller ($R \sim$ 0.417 and 0.450 \rsun) with correspondingly lower luminosity and \teff\ (black data points on Fig. \ref{fig:KrausEB}), and thus provide an independent test to UScoCTIO 5 in a lower mass, but presumably coeval, system. The best-fit spot covering fractions found by \citet{macdonald2017}, respectively for the two components, are 0.22 $\pm$ 0.07 and 0.36 $\pm$ 0.07. Assuming an age of 10~Myr, we find \fspot\ values of 0.22 and 0.43 for the two stars respectively best match the measured \teff\ at the known mass, again matching well the results of \citet{macdonald2010}.

The excellent agreement between our findings and \citet{macdonald2017} for both a mid- and late-M dwarf binary suggests that the precise details of magnetic modelling have only a small effect on the results. Their starspot treatment is fundamentally similar to ours, both having been based off the pioneering work of \citet{spruit1986}, so it is perhaps not surprising. However, it is noteworthy that our results match their despite building upon different lineages of stellar evolutionary models and independent implementations of magnetic activity, suggesting that predictions for the impact of starspots on fundamental parameters are robust to the treatment of stellar structure.

This example demonstrates the power of magnetic models in reproducing the physical properties of young stars. As more pre-MS EBs are characterized in the coming years, magnetic models will be a crucial tool for establishing the young star mass-radius relation, and inversely the physical parameters of young stars will be an invaluable calibrator of next-generation stellar models.

We close by noting an interesting tension in the field: another avenue for testing the radii of M dwarfs is interferometry of field stars \citep{boyajian2012}. The total L and \teff\ can be reconstructed from a combination of spectroscopy and photometry, permitting a test of the L-R-\teff concordance of low mass stars \citep{mann2015}. The results can then be mapped to the mass-radius plane either through models or through empirical M-L relationships in eclipsing binaries. The resulting pattern is different, with a zero-point radius offset in inactive stars and no clear trend with stellar activity.  However, we note that the active and inactive star temperature and luminosity fits were derived using the same relationships; if activity impacts color-\teff\ and flux distributions significantly, this could induce a differential effect not captured in the method, and possibly explain the apparent absence of an activity trend in the interferometric data. Although out of the current scope, the models here could be used to potentially test this effect.

\subsection{Deriving masses and ages of pre-main sequence stars}

The masses and ages of young stars are routinely inferred by comparing their H-R diagram location to stellar isochrones and evolutionary tracks. Because of this, one  consequence of the structural effects of starspots is that the inferred masses and ages of young stars will be altered, as the magnetic tracks following a different evolutionary path \citep[e.g.][]{macdonald2010,jackson2014,somers2015a}. In this section, we assess at first order the magnitude and sign of this change predicted by the SPOTS models.

One important piece of context for this experiment is the curious mass-dependent spread in the inferred ages of pre-main sequence stars that has been found in a number of young open clusters. When measured with stellar isochrones, young M dwarfs are found to be younger than the F stars within the same cluster by up to a factor of two. Example clusters where this phenomenon has been noted include the Upper Scorpius association -- both through studies in the H-R diagram \citet{pecaut2012,feiden2016} and looking at M-F binaries \citep{asensio-torres2019} -- and the $\beta$ Pictoris moving group \citep{malo2014}, among others (see references in \citealt{herczeg2015}). Some authors \citep[e.g.][]{somers2015a,feiden2016} have argued that this discrepancy results from magnetic inflation perturbing the H-R diagram locations of young cool stars, thus producing invalid age determinations from standard stellar models. 

To test whether spot effects could produce such an age discrepancy, we derive masses and ages for a series of Upper Scorpius stars using the SPOTS models. We start from the catalog of Upper Sco stars analyzed in a series of papers about K2 light curves \citep{somers2017b,rebull2018}. \citet{somers2017b} derived temperatures for these stars using photometry, solving simultaneously for reddening -- we adopt these values. Next, we collect Gaia DR2 parallaxes for our sample, discarding stars with distances substantially different from the association mean, and deriving luminosities using the $K_S$ magnitudes and bolometric corrections from the Stellar Interpolation Package for PYthon (SIPPY), a library of interpolation routines for extinction and stellar parameter estimation (Cao et al., in prep). Here, SIPPY takes effective temperature, extinction, and photometric measurements as inputs, and assuming a spot filling factor, returns luminosity, stellar age, and mass parameters as outputs from our SPOTS tracks using a self-consistent spotted perturbation of the \citet{pecaut2013} dwarf color table (Table 5). 

\begin{figure*}
\centering
\includegraphics[width=0.9\linewidth]{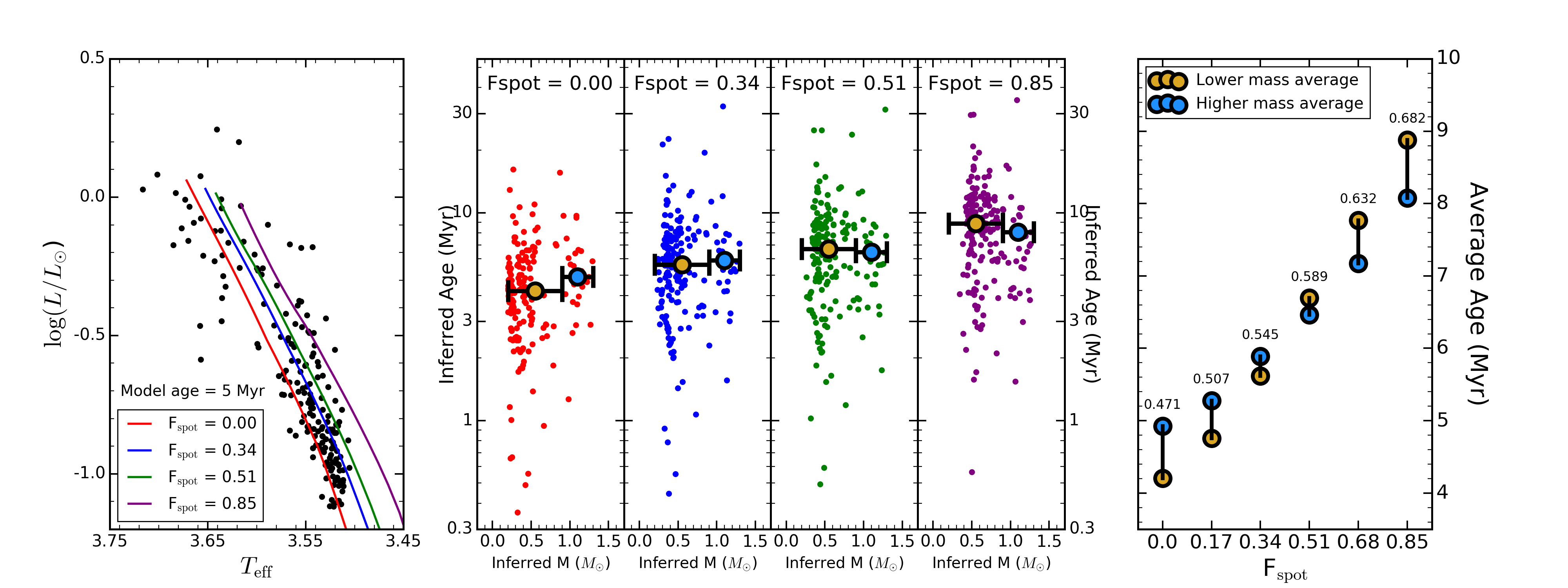}
\caption{A test of the influence of starspots on the derived masses and ages of pre-main sequence stars. {\it Left:} H-R diagram including stars from the Upper Sco association. Also plotted are 5~Myr isochrones for four different starspot covering fractions, from 0.00 to 0.85. {\it Center:} The masses and ages derived for each Upper Sco star from models with different starspot fractions. The yellow and cyan circles reflect the average age above and below 0.9\msun. {\it Right:} The average age of the higher and lower mass ranges for every \fspot\ value in the SPOTS models. At \fspot\ = 0, the higher mass stars have older measured ages. As \fspot\ increases, the average derived ages increase both for higher and lower masses, but the lower mass average overtakes and surpasses the higher mass average. The numbers above each pair of points is the average derived mass of all Upper Sco stars, demonstrating that this metric too increases with increasing \fspot.
\label{fig:UpperSco}
}
\end{figure*}

We plot the catalog along with 3~Myr models with four different starspot properties in the left panel of Fig. \ref{fig:UpperSco}. From red to purple, the models have increasingly larger spot filling factors, and consequently are shifted to cooler temperatures. There exists significant scatter in the distribution of stars, a consequence of a combination of several observational and intrinsic complications \citep[see e.g.][]{hartmann2001,fang2017} along with a possible modest age spread. However, it can be seen from this figure why the inferred ages of stars change when using magnetic models -- by shifting an isochrone to lower temperatures, the models require more contraction time to reach a given spot on the H-R diagram.

Using our SPOTS mass tracks, we do a four-point Lagrangian interpolation across age to produce a regular grid on starspot $\alpha$, age, and mass. Next, we perform a trilinear interpolation over our evolutionary tracks at a fixed spot temperature contrast, outputting luminosity and temperature as a function of age, mass, and spot filling factor. Propagating our luminosity and temperature errors from our spectral types, extinctions, and photometries, we perform a 2D Bayesian analysis in age and mass with uniform age and mass priors. We then marginalize on our age and mass probabilities for each star in order to obtain uncertainties on age and mass. We perform this exercise for each spot filling factor in the SPOTS models (0.00-0.85), leaving us with a set of derived masses and ages of each Upper Sco star in our catalog for each fill factor.

We present the resulting masses and ages for four different spot fill factors in the four central panels of Fig. \ref{fig:UpperSco}. From left to right, we see a general trend of increasing average age with stronger spot properties. There is a mass dependence to this trend as well -- in yellow and cyan, we plot the average ages of the lower and higher mass stars, split at 0.9\msun. The lower-mass end appear younger in the non-spot models, but the discrepancy in ages decreases, and ultimately reverses when you derive the ages with increasingly spotted models. This is shown most clearly in the right-most panel of Fig. \ref{fig:UpperSco}, where we plot the ages of the lower and higher mass samples as a function of \fspot. Both increase towards higher \fspot, but the lower-mass average starts at a lower age and overtakes the higher-mass bin around \fspot\ = 0.5. What this suggests is that starspot activity at the 50\% level can explain the mass discrepancy seen between M- and F-stars in Upper Sco.

Finally, above each set of points we plot the average derived mass of the full sample, in units of the Sun. The average increases monotonically towards high spot covering fractions, reflecting the right-ward movement of individual model tracks in the H-R diagram with increasing spot coverage. This shows that magnetic inflation may induce a bias in the measured initial mass function of young clusters, and must be treated carefully in IMF studies, as pointed out by \citet{stassun2014b}.

Our finding of a slower pre-main sequence among magnetically active stars mirrors the results of previous works \citep[e.g.][]{macdonald2010,somers2014,jackson2014,feiden2016}. It is interesting that these authors reached similar result despite the varying treatments of magnetic activity -- these works variously model magnetism as a reduced mixing length, internal magnetic fields inhibiting convection, and surface starspots. That results of similar order are found in each case suggests a modeling degeneracy between the major families of active-star models, at least as far as the Hayashi contraction rate is concerned. Perhaps then the best path forward for distinguishing between these scenarios with data lies in analyzing color and temperature perturbations in active stars.

A comprehensive analysis of Upper Sco and other moving groups with our magnetic models is outside of the scope of this section, but we have demonstrated:

\begin{itemize}

\item[\textbullet] Ages derived from Upper Sco with standard models imply a mass-dependent age gradient. 

\item[\textbullet] The average derived age of stars increases when using inflated models, suggesting that standard models systematically under-estimate the ages of magnetically-active young stars.

\item[\textbullet] The rate of change in derived age is mass dependent, such that an \fspot\ value exists where the higher and lower mass ends agree in average age.

\item[\textbullet] The derived masses of stars increase when using more inflated models.

\end{itemize}

As we do not yet know the true correlation between mass and average starspot properties on the pre-main sequence, there is no particular reason to think that constant magnetic properties as a function of mass, or even between different stars of equal mass, provide the best representation of the population. Future studies should attempt to directly compare stars with known starspot properties to their respective magnetic models in order to test the structural predictions of the SPOTS models.

\begin{figure*}
\centering
\includegraphics[width=0.9\linewidth]{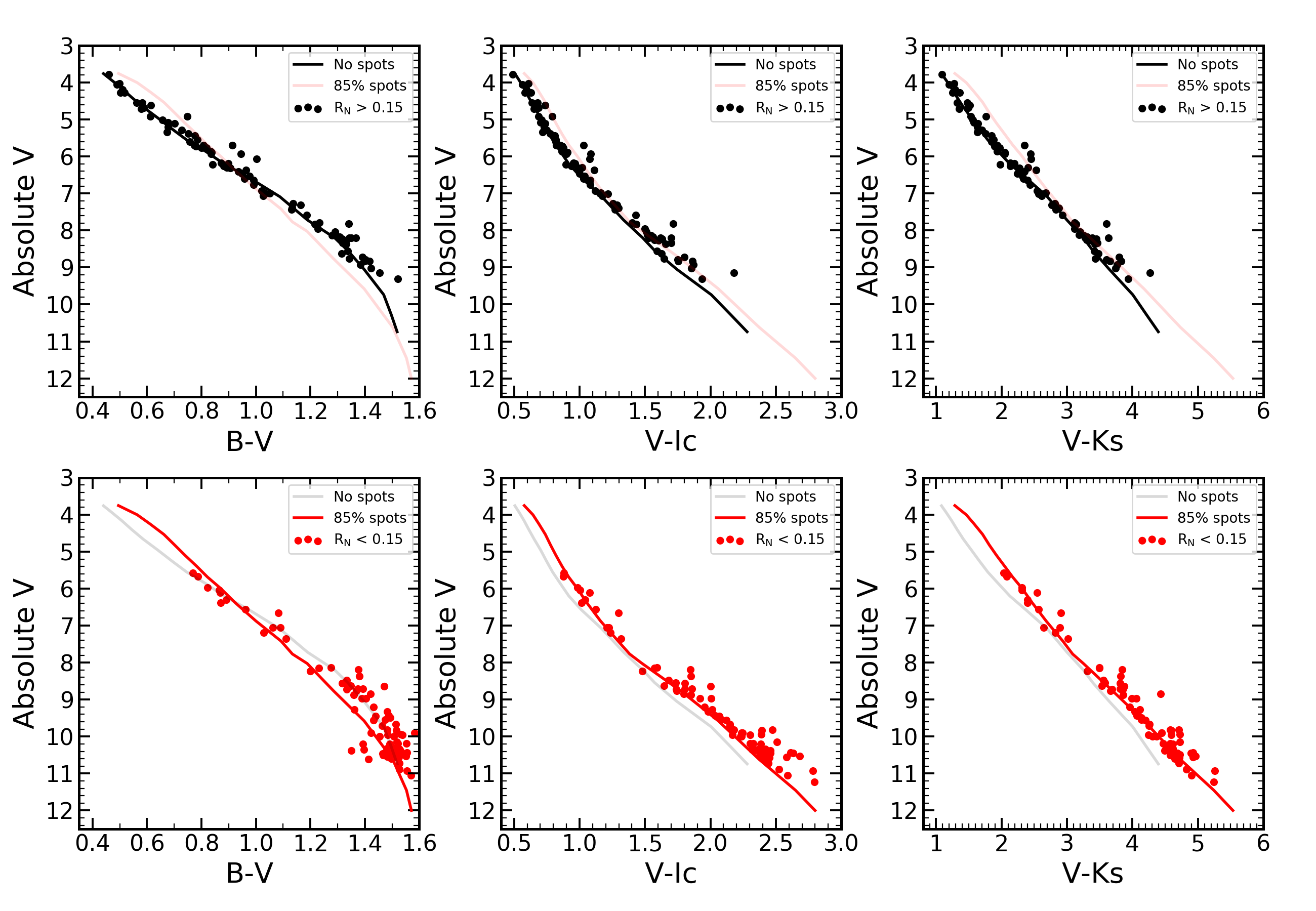}
\caption{A comparison of colors from the SPOTS models to low-mass stars in the Pleiades as a function of rotation. {\it Top:} Black and red lines represent the SPOTS models of spot-less and \fspot\ = 0.85 at 120~Myr in three different color bands. Black points are Pleiads with a Rossby number greater than 0.15, indicating slower rotating. Agreement with the spot-less models is overall very good, but not with the spotted models. {\it Bottom:} Same as top, except for stars with Rossby number less than 0.15, indicating rapid rotation. In the central and right panels, the \fspot\ = 0.85 model predicts the data better than the spot-less model, though it is less clear in the left-most panel.
\label{fig:Pleiades}
}
\end{figure*}

\subsection{The Colors of Active Stars}\label{sec:activestarcolors}

There are two mechanisms through which our models alter the color scale of stars: radius inflation, which perturbs the \teff\ and luminosity, and starspots, which alter the shape of SED. In this section we present two comparisons of our models with active stars, one with stars in a young open cluster and one with calibrated color relations from pre-main sequence stars, in order to assess the validity of our model predictions. We note that comprehensive comparisons are out-of-scope for this paper, and instead seek illustrative examples of model efficacy. 

\subsubsection{The K-dwarfs of the Pleiades}

For decades, it has been known that the K dwarfs in the $\sim 120$~Myr old Pleiades open cluster have abnormal colors compared to older open clusters and field stars. The sense of the abnormality is that in bluer color bands, such as $B-V$, the Pleiads are ``too blue'' compared to a single-star isochrone. However in redder bands, such as $V-K_S$, they are ``too red'' compared to a single-star isochrone. This discovery dates to at least \citet{herbig1962}, and has been suggested by \citet{stauffer2003} to be a manifestation of intense starspot and chromosphere activity. This hypothesis is supported by the work of \citet{kamai2014}: these authors found a positive correlation between rotation rate and color abnormality, suggesting that the most rapidly rotating, and thus perhaps the most heavily spotted stars, show the greatest abnormalities. A rotation-color correlation was later demonstrated for the stars less massive than spectral-type K by \citet{covey2016}. Moreover, \citet{somers2017a} found a connection between rotation rate and radius inflation within this cluster, signifying that magnetic activity is likely influencing the structure of these K-dwarfs.

Given the apparent influence of starspots on the colors of the Pleiades, this cluster provides an interesting comparison with our starspot models. In Fig. \ref{fig:Pleiades}, we show $B-V$, $V-I_C$, and $V-K_S$ measurements of low-mass Pleiades stars from \citet{kamai2014}, relative to absolute V magnitude -- calculated using d = 136~pc from Gaia DR1 \citet{gaiadr1_2016}. In the top and bottom rows, we have separated the slower (black points) and faster (red points) rotating stars by adopting a Rossby number\footnote{The Rossby number is defined as the ratio of the rotation period to the convective overturn timescale \citep[e.g.][]{pizzolato2003}} cut off. Because the Rossby number scales with the convective properties of stars, it provides a more natural comparison of the magnetic properties of stars of difference mass than just the rotation period. 

\begin{figure*}
\centering
\includegraphics[width=0.9\linewidth]{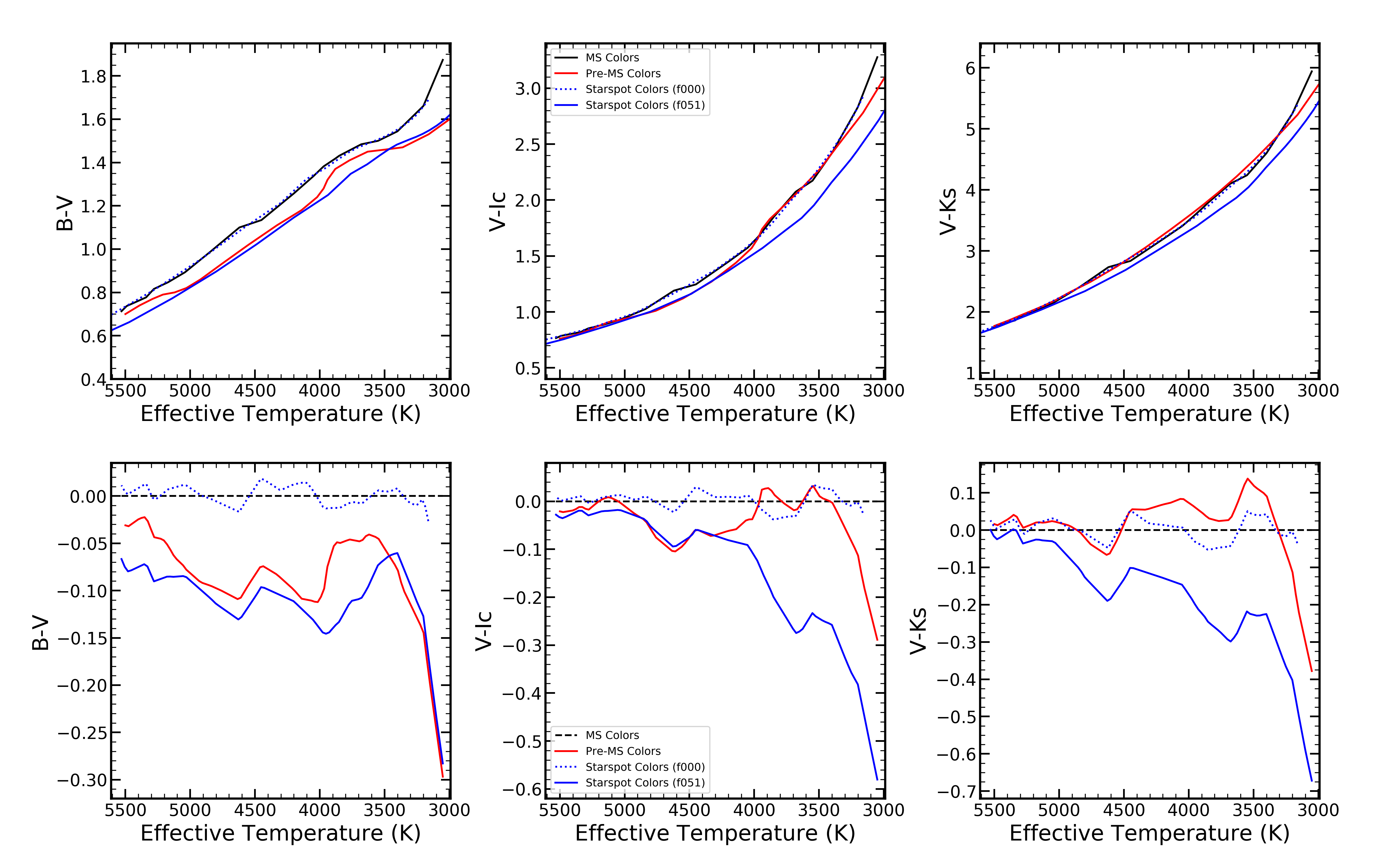}
\caption{A comparison of colors from the SPOTS models to the pre-main sequence colors of \citet{pecaut2013}. {\it Top:} A comparison of three different colors as a function of \teff. The black and red lines shows the main sequence and pre-main sequence color calibrations from PM13, respectively. The dotted and solid blue lines are colors from our spot-less and \fspot\ = 0.51 models, respectively. The dotted line agrees well with the black line because the PM13 main sequence colors were the source of our calibration, so our spot-less model will simply reproduce that relation. The solid blue line shows the predictions of the spot colors. In the first panel, they show excellent agreement with the pre-main sequence colors, suggesting that the offset between PMS and MS could be a product of starspots. The agreement is not as good at the cool end of the center panel, or at any temperature in the right panel. {\it Bottom:} Detrending the top panels relative to the MS Colors line. The qualitative agreement at higher temperatures, and the divergence at lower temperatures for $V-I_C$ and $V-K_S$ is readily apparent.
\label{fig:PMScolors}
}
\end{figure*}

Both the rapid and slow stars are compared to a no-spot model (black) and an 85\% spot model (red), adopting our two-component color transformation scheme. Looking first to the top row, we find that the no-spot stellar isochrone matches extremely well the locus of slowly-rotating Pleiads in the whole mass range. The starspot model deviates from this locus in a color-dependent fashion. In $V-I_C$ and $V-K_S$, the spot models are redder at all absolute V-band magnitudes, providing a poorer fit to the slowly rotating data, except perhaps in the case of binaries. In the $B-V$ panel, the spot model is redder for stars brighter than $V \sim 6.5$, and bluer for dimmer stars. Consequently, the starspot models are a poorer fit to the slowly rotating $B-V$ data as well.

Looking now to the faster rotators in the bottom row, we no longer find excellent agreement between the single-star locus and the no-spot models in $V-I_C$ and $V-K_S$. For the entire data set, but particularly for the portion dimmer than $V \sim 8$, the single-star locus is noticeably redder than the no-spot model by perhaps 0.2-0.4~mag. Remarkably, the left edge of the stellar distribution now agrees extremely well with the predictions of the starspot model. This suggests that rapid rotation is distorting the colors of these objects, perhaps as a combination of magnetically-induced radius inflation (hence lower average surface temperature) and the direct impact of cool spots on the SED. 

The picture is less clear in $B-V$ -- the majority of the rapid rotators still appear closer to the no-spot model below $V \sim 9$, though the width of the main sequence has broadened considerably -- this has been suggested as the result of a range of magnetic activity levels and spot activity \citep{guo2018}. However, a number of stars are clearly much bluer that the primary locus, a potential signature of magnetic activity. It is likely that our color transformations are not as accurate for colors containing B-band light because we have not modeled chromospheric emission arising from intense magnetic activity, which may contribute significantly to blue color bands \citep[e.g.][]{stauffer2003}. Including a model of chromospheric emission is far beyond the scope of this paper, as it would require at least a three-component temperature fit.

\citet{jackson2014} performed a similar experiment, considering the influence of starspots on synthetic $B-V$ and $V-K_S$ colors at 120~Myr. Their $B-V$ findings are similar to ours, producing bluer colors for K-dwarfs (in accordance with data). However this blueward shift is not uniform in our models -- among G-dwarfs, our models suggest that spotted stars may be redder in $B-V$. The situation also differs in $V-K_S$ colors, where \citet{jackson2014} find a blueward shift driven by stars and we find a redward shift. The reason for the discrepancy is unclear, but may relate to changes in temperature of the un-spotted surface in our models -- $T_{\rm hot}$ rises marginally in the presence of starspots. Subtler effects like different choices of bolometric correction and color scale also may play a role. The observations of \citet{kamai2014} appear to support a redder $V-K_S$ color for low Rossby-number stars than expected from calibrations on in-active stars, as have some other findings \citet[e.g.][]{covey2016}, however this could also result from magnetic inhibition of convection with little spot coverage \citet{jackson2014}.

From this simple comparison, we have shown that our models can accurately reproduce the distorted SEDs of stars in redder bands, and that perhaps more components are needed to reproduce the bluer SED. Another possibility is that our adopted spot temperature contrast (0.8) should be adjusted to precisely account for the effects on the SED in the blue. Future work should consider the source of the discrepancy in $B-V$.

\subsubsection{Pre-main sequence colors from Pecaut \& Mamajek (2013)}

We can also check our model colors by comparing to calibrated color-\teff\ relations for pre-MS stars. It has been shown in several works that the color scale on the pre-MS differs in certain interesting ways from the main sequence color scale. Possible sources of the differing colors of young stars include the presence of cool spots, intense chromospheric/coronal activity, and their lower intrinsic surface gravities. We can thus pose an interesting question: how much of this discrepancy can be accounted for by the cool spots? The color scales used to derive synthetic photometry for our models drew from main-sequence sources, so the resulting two-temperature colors are largely independent of the confounding effects of low surface gravity and chromospheric/coronal emission on the pre-MS. 

In addition to their main-sequence color tables, PM13 produced a widely-used set of colors for pre-MS objects, ages 5--30~Myr. We compare in Fig. \ref{fig:PMScolors} our model colors to the pre-MS and MS color scale, relative to \teff, from these authors. We choose the same three color bands as in Fig. \ref{fig:Pleiades} for consistency. In black and red, we plot the MS and pre-MS color-\teff\ relations from PM13. The dotted blue line shows the resulting colors from our no-spot models at 30 Myr; naturally, these track the main-sequence colors that their photometry was calibrated on. Finally, in solid blue we show our 51\% starspot models. The top row shows absolute colors versus \teff, and the bottom shows colors detrended with respect to the PM13 main sequence colors.

Looking first at $B-V$ we find that the pre-MS colors are systematically bluer at fixed \teff\ by approximately 0.1-0.15~mag. Remarkably, our starspot models line up right on top of the pre-MS colors throughout the entire \teff\ range. This is confirmed in the detrended plot on bottom. The proximate reason is the slightly increased temperature of the ambient, non-spot regions compared to a star of equivalent mass and age but without spots. This higher temperature reaches farther down the Wien-tail of the stellar spectrum and increases the total B-band light, thus making the models bluer. 

However, the models do not line up with the young star colors as well in the other two color bands. For $V-I_C$, the spot models track the empirical colors well hotter than $\sim 4200$~K, when the pre-MS colors begin to agree well with the MS colors, and our models remain bluer. And in $V-K_S$, our models diverge from the empirical relations quite early -- they remain bluer than the empirical pre-MS relation at all temperatures below $\sim 5000$~K, whereas the MS colors agree well with the pre-MS colors. This uneven agreement may be a consequence of gravity-dependent color deformations \citep[e.g.][]{gullbring1998,dario2010,pecaut2013}, resulting from differing absorption line equivalent widths in a lower-pressure photosphere -- it is unlikely that the main sequence empirical tables we have used to generate our synthetic colors would capture this effect.

\subsection{Sub-subgiants}

\begin{figure*}
\centering
\includegraphics[width=0.9\linewidth]{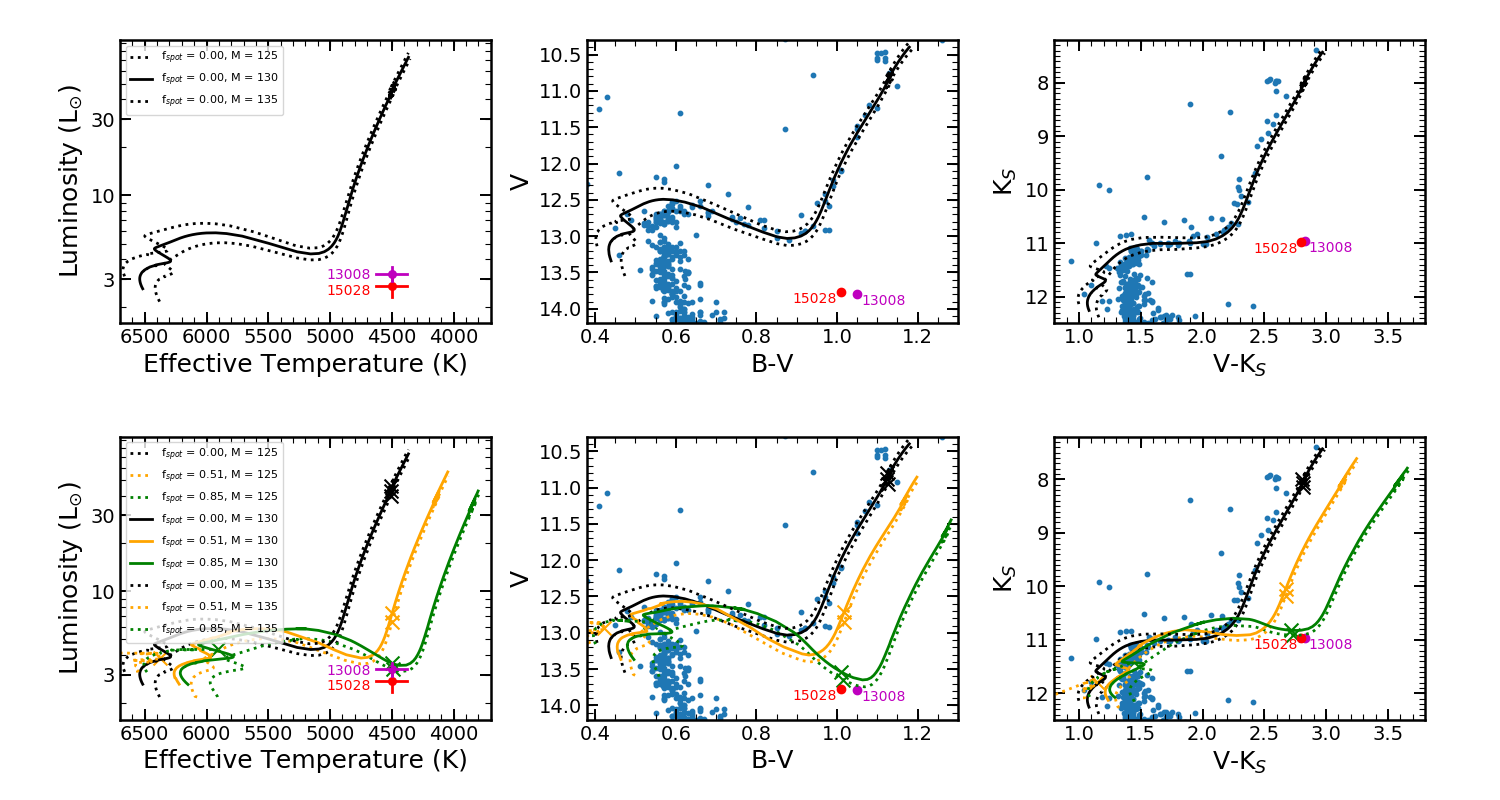}
\caption{A comparison of the SPOTS models to the sub-subgiants of M67. {\it Top Left:} Spot-less models from 1.25-1.35\msun, the approximate turn off mass of M67, compared to two SSGs from the cluster. The models cannot reproduce their location at all. {\it Top Center/Right:} Same as top left, but in two different CMDs. We also show a catalog of M67 members as blue points, demonstrating that our models predict the post-MS of normal stars accurately. {\it Bottom:} Same as top, but now including \fspot\ = 0.51 (yellow) and 0.85 (green) SPOTS models. The models cool and reduce their luminosity enough to predict the locations of the SSGs in all three panels simultaneously.
\label{fig:M67ssgs}
}
\end{figure*}

Sub-subgiants (SSGs) are a peculiar class of objects which reside below the traditional subgiant locus on both the Hertzprung-Russell and color-magnitude diagrams. They have been known about for some time \citep{belloni1998}, but have continued to defy explanation. In a recent series of papers, \citet{geller2017a,leiner2017,geller2017b} provided a comprehensive study of literature SSGs and explored several possible formation mechanisms. These authors concluded that the most promising explanation for SSGs is that they are post-MS stars forced to a rapid rate of rotation by a tidally-locked binary companion. This fast rotation, persisting as the surface convection zone deepens due to structural evolution on the subgiant branch, produces strong magnetic activity, dramatically cooling the star and suppressing its luminosity. This hypothesis is supported by several properties of SSGs: their frequent binary nature, their large photometric modulation amplitudes (and hence intense starspot coverage), their strong X-ray, H-alpha, and Ca H\&K emission, and their enhanced rotation rates relative to typical sub-giants. The conclusion was further bolstered by MESA models presented by \citet{leiner2017}, who included magnetic inflation by reducing the mixing-length ratio as was done by \citet{chabrier2007}. These authors found that strong convective suppression can qualitatively reproduce the H-R diagram positions of two SSGs in the old open cluster M67. 

With our SPOTS models, we can build off of this initial modeling exercise and test not only the ability of magnetic models to reproduce the H-R diagram location, but also whether our two-temperature color models can predict the SSG locale in various color bands. In Fig. \ref{fig:M67ssgs}, we compare the post-MS tracks of 1.25, 1.30, and 1.35\msun\ models -- the approximate range of estimates for the turn-off mass of the cluster \citep{leiner2017} -- to M67. Blue points are identified M67 members, and the red and purple points are the SSGs WOCS15028 and WOCS13008, studied by \citet{leiner2017}. On the top panel we show our non-spot models, demonstrating their accurate prediction of the cluster turn-off in the two CMDs, and also that the SSGs are in an anomalous location without a traditional stellar model explanation.

On the bottom row of Fig. \ref{fig:M67ssgs} we include the predictions from our 51\% and 85\% spot models. The influence of magnetic starspots on the post-MS is similar to the pre-MS effect: stars are cooler and less luminous, so that the local luminosity minimum on the subgiant branch is approximately 500-600~K cooler than the standard model for 85\% spots. This degree of spot activity allows the model to reproduce the SSGs, similar to the MESA models run by \citet{leiner2017}. We also compare our two-temperature colors to the CMD locations of the SSGs, finding remarkably agreement in both $B-V$ and $V-K_S$. Moreover, the agreement with each model all occurs at precisely the same epoch -- the green "x" in each panel reflects a single age of the stellar model, and reflects in all three panels the point of closest approach between the model predictions and the SSGs themselves.

This exercise demonstrates that magnetic models with strong surface fields are capable of reproducing both the fundamental parameters and colors of sub-sub giants. The extremity of this class of objects makes them a valuable test-bed for magnetic models -- they deserve significantly more attention in future publications.

\section{Summary and Conclusions}

In this paper we have presented the Stellar Parameters of Tracks with Starspots (SPOTS) models, a new suite of stellar evolution tracks and isochrones which incorporate the treatment of surface activity and starspots introduced in \citetalias{somers2015a}. Our models consider both the inhibition of convective energy transport due to magnetic field lines near the surface layers, and the influence of cool surface regions on the boundary conditions at the photosphere. These effects have important structural consequences, tracked in a consistent manner by our calculations. We also produce colors for our models using the empirically-calibrated main sequence tables of \citet{pecaut2013}, accounting for the fact that different spectra are emitted from the hot and cool regions of two-temperature spotted stars.

With these models, we briefly consider a number of outstanding puzzles in the study of active stars, including the anomalously large radii of short-period eclipsing binaries, the apparent mass trends in the ages of stars in young associations and clusters, the peculiar colors of active K-dwarfs in the Pleiades, the inconsistencies between pre-main sequence and main sequence Teff-color relations, and the existence of the mysterious class of objects called sub-subgiants. In each case, we find that intense magnetic activity perturbing the structure of stars offers a plausible explanation for these phenomena. Future work should undertake more detailed investigations of each of these particular issues, as we have only endeavored to make plausibility arguments. 

One motivation behind constructing these models is to test whether the known offsets between pre-main sequence and main sequence colors results from distortions in the SEDs of heavily spotted stars, due to the multi-temperature nature of their photospheres. Two-temperature spectral and photometric models have been explored in the past \citep[e.g.][]{stauffer2003,jackson2009,macdonald2013,fang2017}, in some cases simultaneously treating the influence of activity of stellar structure, and we hope that by releasing a set of evolutionary tracks to the community that active stellar colors may become more widely employed. We find in $\S$\ref{sec:activestarcolors} that our models can partially explain the discrepancy -- our models predict very well the Teff-$B-V$ correlation of pre-main sequence stars, but not for redder colors such as $V-K_S$. This partial success may indicate that our empirical colors are incomplete representations of the long-wavelength behavior of starspots, or that a two-temperature model is insufficiently complex to predict the full SED of heavily spotted stars. This is an interesting tension to consider for improving the accuracy of our colors going forward.

While the influence of activity on the early evolution of pre-main sequence stars remains a vexing and unsolved problem, a resolution is most likely to come from synergies between theoretical modeling and empirical studies. One form this connection can take is the application of empirical corrections to the colors of normal, non-magnetic evolutionary calculations \citep{stassun2012,bell2012,chen2014}. This method involves comparing a series of active stars to the predictions of evolutionary models, calculating the offset, and applying that offset to different stars at various ages. These models produce good results (by construction) and are useful for classifying young stars, but because of the \textit{ad hoc} color perturbations they are not completely physically self-consistent.

Magnetic evolutionary models, such as SPOTS, provide an alternative method for exploring the evolution of young stars which are physically self-consistent and independent of data. Our models also provide directly-testable predictions of the magnitude of magnetic effects, permitting a more granular approach to classifying active stars. However, a limitation of our models it that we only treat two-temperature stellar surfaces whereas active photospheres can be significantly more complicated \citep[e.g.][]{roettenbacher2016,montet2017}. Empirical calibrations capture this effect by default, but theoretical models must be expanded to create this new functionality. Future iterations of magnetic models should focus on refining their models of photospheric emissions.

\section{Acknowledgements}

We thank John Stauffer for a helpful suggestion on how best to determine \teff\-color relations for the new Gaia data, Eric Mamajek for advice on using the PM13 main sequence color tables, Gregory Feiden for kindly providing the newest version of his evolutionary models, and Keivan Stassun for helpful advice along the way. GS acknowledges the support of the Vanderbilt Office of the Provost through the Vanderbilt Initiative in Data-intensive Astrophysics (VIDA). This work has made use of data from the European Space Agency (ESA) mission Gaia (https://www.cosmos.esa.int/gaia), processed by the Gaia Data Processing and Analysis Consortium (DPAC, https://www.cosmos.esa.int/web/gaia/dpac/consortium). Funding for the DPAC has been provided by national institutions, in particular the institutions participating in the Gaia Multilateral Agreement. GS acknowledges and thanks all of the wonderful mentors, collaborators, and friends that he met during his astronomy journey, in particular Keivan Stassun and Marc Pinsonneault, the two most influentual mentors on my development as a professional and scientist. 

\bibliographystyle{aasjournal}
\bibliography{biblio}

\begin{thebibliography}{}
\expandafter\ifx\csname natexlab\endcsname\relax\def\natexlab#1{#1}\fi

\bibitem[{{Adelberger} {et~al.}(2011){Adelberger}, {Garc{\'{\i}}a},
  {Robertson}, {Snover}, {Balantekin}, {Heeger}, {Ramsey-Musolf}, {Bemmerer},
  {Junghans}, {Bertulani}, {Chen}, {Costantini}, {Prati}, {Couder},
  {Uberseder}, {Wiescher}, {Cyburt}, {Davids}, {Freedman}, {Gai}, {Gazit},
  {Gialanella}, {Imbriani}, {Greife}, {Hass}, {Haxton}, {Itahashi}, {Kubodera},
  {Langanke}, {Leitner}, {Leitner}, {Vetter}, {Winslow}, {Marcucci},
  {Motobayashi}, {Mukhamedzhanov}, {Tribble}, {Nollett}, {Nunes}, {Park},
  {Parker}, {Schiavilla}, {Simpson}, {Spitaleri}, {Strieder}, {Trautvetter},
  {Suemmerer}, \& {Typel}}]{adelberger2011}
{Adelberger}, E.~G., {Garc{\'{\i}}a}, A., {Robertson}, R.~G.~H., {et~al.} 2011,
  Reviews of Modern Physics, 83, 195

\bibitem[{{Allard} {et~al.}(1997){Allard}, {Hauschildt}, {Alexander}, \&
  {Starrfield}}]{allard1997}
{Allard}, F., {Hauschildt}, P.~H., {Alexander}, D.~R., \& {Starrfield}, S.
  1997, \araa, 35, 137

\bibitem[{{Asensio-Torres} {et~al.}(2019){Asensio-Torres}, {Currie}, {Janson},
  {Desidera}, {Kuzuhara}, {Hodapp}, {Brandt}, {Guyon}, {Lozi}, \&
  {Groff}}]{asensio-torres2019}
{Asensio-Torres}, R., {Currie}, T., {Janson}, M., {et~al.} 2019, \aap, 622, A42

\bibitem[{{Asplund} {et~al.}(2005){Asplund}, {Grevesse}, \&
  {Sauval}}]{asplund2005}
{Asplund}, M., {Grevesse}, N., \& {Sauval}, A.~J. 2005, in Astronomical Society
  of the Pacific Conference Series, Vol. 336, Cosmic Abundances as Records of
  Stellar Evolution and Nucleosynthesis, ed. I.~{Barnes}, Thomas~G. \& F.~N.
  {Bash}, 25

\bibitem[{{Azulay} {et~al.}(2017){Azulay}, {Guirado}, {Marcaide},
  {Mart{\'\i}-Vidal}, {Ros}, {Tognelli}, {Hormuth}, \& {Ortiz}}]{azulay2017}
{Azulay}, R., {Guirado}, J.~C., {Marcaide}, J.~M., {et~al.} 2017, \aap, 602,
  A57

\bibitem[{{Baraffe} {et~al.}(2009){Baraffe}, {Chabrier}, \&
  {Gallardo}}]{baraffe2009}
{Baraffe}, I., {Chabrier}, G., \& {Gallardo}, J. 2009, \apjl, 702, L27

\bibitem[{{Bell} {et~al.}(2012){Bell}, {Naylor}, {Mayne}, {Jeffries}, \&
  {Littlefair}}]{bell2012}
{Bell}, C. P.~M., {Naylor}, T., {Mayne}, N.~J., {Jeffries}, R.~D., \&
  {Littlefair}, S.~P. 2012, \mnras, 424, 3178

\bibitem[{{Belloni} {et~al.}(1998){Belloni}, {Verbunt}, \&
  {Mathieu}}]{belloni1998}
{Belloni}, T., {Verbunt}, F., \& {Mathieu}, R.~D. 1998, \aap, 339, 431

\bibitem[{{Berdyugina}(2005)}]{berdyugina2005}
{Berdyugina}, S.~V. 2005, Living Reviews in Solar Physics, 2, 8

\bibitem[{{Boyajian} {et~al.}(2012){Boyajian}, {von Braun}, {van Belle},
  {McAlister}, {ten Brummelaar}, {Kane}, {Muirhead}, {Jones}, {White},
  {Schaefer}, {Ciardi}, {Henry}, {L{\'o}pez-Morales}, {Ridgway}, {Gies}, {Jao},
  {Rojas-Ayala}, {Parks}, {Sturmann}, {Sturmann}, {Turner}, {Farrington},
  {Goldfinger}, \& {Berger}}]{boyajian2012}
{Boyajian}, T.~S., {von Braun}, K., {van Belle}, G., {et~al.} 2012, \apj, 757,
  112

\bibitem[{{Browning}(2008)}]{browning2008}
{Browning}, M.~K. 2008, \apj, 676, 1262

\bibitem[{{Campbell}(1984)}]{campbell1984}
{Campbell}, B. 1984, \apj, 283, 209

\bibitem[{{Castelli} \& {Kurucz}(2004)}]{castelli2004}
{Castelli}, F., \& {Kurucz}, R.~L. 2004, ArXiv Astrophysics e-prints,
  astro-ph/0405087

\bibitem[{{Chabrier} {et~al.}(2007){Chabrier}, {Gallardo}, \&
  {Baraffe}}]{chabrier2007}
{Chabrier}, G., {Gallardo}, J., \& {Baraffe}, I. 2007, \aap, 472, L17

\bibitem[{{Chen} {et~al.}(2014){Chen}, {Girardi}, {Bressan}, {Marigo},
  {Barbieri}, \& {Kong}}]{chen2014}
{Chen}, Y., {Girardi}, L., {Bressan}, A., {et~al.} 2014, \mnras, 444, 2525

\bibitem[{{Choi} {et~al.}(2016){Choi}, {Dotter}, {Conroy}, {Cantiello},
  {Paxton}, \& {Johnson}}]{choi2016}
{Choi}, J., {Dotter}, A., {Conroy}, C., {et~al.} 2016, \apj, 823, 102

\bibitem[{{Covey} {et~al.}(2016){Covey}, {Ag{\"u}eros}, {Law}, {Liu}, {Ahmadi},
  {Laher}, {Levitan}, {Sesar}, \& {Surace}}]{covey2016}
{Covey}, K.~R., {Ag{\"u}eros}, M.~A., {Law}, N.~M., {et~al.} 2016, \apj, 822,
  81

\bibitem[{{Da Rio} {et~al.}(2010){Da Rio}, {Robberto}, {Soderblom}, {Panagia},
  {Hillenbrand}, {Palla}, \& {Stassun}}]{dario2010}
{Da Rio}, N., {Robberto}, M., {Soderblom}, D.~R., {et~al.} 2010, \apj, 722,
  1092

\bibitem[{{David} {et~al.}(2016){David}, {Hillenbrand}, {Cody}, {Carpenter}, \&
  {Howard}}]{david2016}
{David}, T.~J., {Hillenbrand}, L.~A., {Cody}, A.~M., {Carpenter}, J.~M., \&
  {Howard}, A.~W. 2016, \apj, 816, 21

\bibitem[{{David} {et~al.}(2019){David}, {Hillenbrand}, {Gillen}, {Cody},
  {Howell}, {Isaacson}, \& {Livingston}}]{david2019}
{David}, T.~J., {Hillenbrand}, L.~A., {Gillen}, E., {et~al.} 2019, \apj, 872,
  161

\bibitem[{{Dotter}(2016)}]{dotter2016}
{Dotter}, A. 2016, \apjs, 222, 8

\bibitem[{{Evans} {et~al.}(2009){Evans}, {Dunham}, {J{\o}rgensen}, {Enoch},
  {Mer{\'\i}n}, {van Dishoeck}, {Alcal{\'a}}, {Myers}, {Stapelfeldt}, {Huard},
  {Allen}, {Harvey}, {van Kempen}, {Blake}, {Koerner}, {Mundy}, {Padgett}, \&
  {Sargent}}]{evans2009}
{Evans}, Neal~J., I., {Dunham}, M.~M., {J{\o}rgensen}, J.~K., {et~al.} 2009,
  \apjs, 181, 321

\bibitem[{{Fang} {et~al.}(2017){Fang}, {Herczeg}, \& {Rizzuto}}]{fang2017}
{Fang}, Q., {Herczeg}, G.~J., \& {Rizzuto}, A. 2017, \apj, 842, 123

\bibitem[{{Feiden}(2016)}]{feiden2016}
{Feiden}, G.~A. 2016, \aap, 593, A99

\bibitem[{{Feiden} \& {Chaboyer}(2012)}]{feiden2012}
{Feiden}, G.~A., \& {Chaboyer}, B. 2012, \apj, 757, 42

\bibitem[{{Feiden} \& {Chaboyer}(2013)}]{feiden2013}
---. 2013, \apj, 779, 183

\bibitem[{{Feiden} \& {Chaboyer}(2014)}]{feiden2014}
---. 2014, \apj, 789, 53

\bibitem[{{Ferguson} {et~al.}(2005){Ferguson}, {Alexander}, {Allard}, {Barman},
  {Bodnarik}, {Hauschildt}, {Heffner-Wong}, \& {Tamanai}}]{ferguson2005}
{Ferguson}, J.~W., {Alexander}, D.~R., {Allard}, F., {et~al.} 2005, \apj, 623,
  585

\bibitem[{{Gaia Collaboration} {et~al.}(2016){Gaia Collaboration}, {Prusti},
  {de Bruijne}, {Brown}, {Vallenari}, {Babusiaux}, {Bailer-Jones}, {Bastian},
  {Biermann}, \& {Evans}}]{gaiadr1_2016}
{Gaia Collaboration}, {Prusti}, T., {de Bruijne}, J.~H.~J., {et~al.} 2016,
  \aap, 595, A1

\bibitem[{{Gaia Collaboration} {et~al.}(2018){Gaia Collaboration}, {Brown},
  {Vallenari}, {Prusti}, {de Bruijne}, {Babusiaux}, {Bailer-Jones}, {Biermann},
  {Evans}, \& {Eyer}}]{gaiadr2_2018}
{Gaia Collaboration}, {Brown}, A.~G.~A., {Vallenari}, A., {et~al.} 2018, \aap,
  616, A1

\bibitem[{{Geller} {et~al.}(2017{\natexlab{a}}){Geller}, {Leiner},
  {Chatterjee}, {Leigh}, {Mathieu}, \& {Sills}}]{geller2017b}
{Geller}, A.~M., {Leiner}, E.~M., {Chatterjee}, S., {et~al.}
  2017{\natexlab{a}}, \apj, 842, 1

\bibitem[{{Geller} {et~al.}(2017{\natexlab{b}}){Geller}, {Leiner}, {Bellini},
  {Gleisinger}, {Haggard}, {Kamann}, {Leigh}, {Mathieu}, {Sills}, \&
  {Watkins}}]{geller2017a}
{Geller}, A.~M., {Leiner}, E.~M., {Bellini}, A., {et~al.} 2017{\natexlab{b}},
  \apj, 840, 66

\bibitem[{{Grevesse} \& {Sauval}(1998)}]{grevesse1998}
{Grevesse}, N., \& {Sauval}, A.~J. 1998, \ssr, 85, 161

\bibitem[{{Gullbring} {et~al.}(1998){Gullbring}, {Hartmann}, {Brice{\~n}o}, \&
  {Calvet}}]{gullbring1998}
{Gullbring}, E., {Hartmann}, L., {Brice{\~n}o}, C., \& {Calvet}, N. 1998, \apj,
  492, 323

\bibitem[{{Gully-Santiago} {et~al.}(2017){Gully-Santiago}, {Herczeg},
  {Czekala}, {Somers}, {Grankin}, {Covey}, {Donati}, {Alencar}, {Hussain}, \&
  {Shappee}}]{gully-santiago2017}
{Gully-Santiago}, M.~A., {Herczeg}, G.~J., {Czekala}, I., {et~al.} 2017, \apj,
  836, 200

\bibitem[{{Guo} {et~al.}(2018){Guo}, {Gully-Santiago}, \& {Herczeg}}]{guo2018}
{Guo}, Z., {Gully-Santiago}, M., \& {Herczeg}, G.~J. 2018, \apj, 868, 143

\bibitem[{{Hartmann}(2001)}]{hartmann2001}
{Hartmann}, L. 2001, \aj, 121, 1030

\bibitem[{{Herbig}(1962)}]{herbig1962}
{Herbig}, G.~H. 1962, \apj, 135, 736

\bibitem[{{Herczeg} \& {Hillenbrand}(2015)}]{herczeg2015}
{Herczeg}, G.~J., \& {Hillenbrand}, L.~A. 2015, \apj, 808, 23

\bibitem[{{Hillenbrand}(1997)}]{hillenbrand1997}
{Hillenbrand}, L.~A. 1997, \aj, 113, 1733

\bibitem[{{Jackson} \& {Jeffries}(2014)}]{jackson2014}
{Jackson}, R.~J., \& {Jeffries}, R.~D. 2014, \mnras, 441, 2111

\bibitem[{{Jackson} {et~al.}(2009){Jackson}, {Jeffries}, \&
  {Maxted}}]{jackson2009}
{Jackson}, R.~J., {Jeffries}, R.~D., \& {Maxted}, P.~F.~L. 2009, \mnras, 399,
  L89

\bibitem[{{Jeffries} {et~al.}(2017){Jeffries}, {Jackson}, {Franciosini},
  {Randich}, {Barrado}, {Frasca}, {Klutsch}, {Lanzafame}, {Prisinzano},
  {Sacco}, {Gilmore}, {Vallenari}, {Alfaro}, {Koposov}, {Pancino}, {Bayo},
  {Casey}, {Costado}, {Damiani}, {Hourihane}, {Lewis}, {Jofre}, {Magrini},
  {Monaco}, {Morbidelli}, {Worley}, {Zaggia}, \& {Zwitter}}]{jeffries2017}
{Jeffries}, R.~D., {Jackson}, R.~J., {Franciosini}, E., {et~al.} 2017, \mnras,
  464, 1456

\bibitem[{{Kamai} {et~al.}(2014){Kamai}, {Vrba}, {Stauffer}, \&
  {Stassun}}]{kamai2014}
{Kamai}, B.~L., {Vrba}, F.~J., {Stauffer}, J.~R., \& {Stassun}, K.~G. 2014,
  \aj, 148, 30

\bibitem[{{Kraus} {et~al.}(2015){Kraus}, {Cody}, {Covey}, {Rizzuto}, {Mann}, \&
  {Ireland}}]{kraus2015}
{Kraus}, A.~L., {Cody}, A.~M., {Covey}, K.~R., {et~al.} 2015, \apj, 807, 3

\bibitem[{{Leiner} {et~al.}(2017){Leiner}, {Mathieu}, \& {Geller}}]{leiner2017}
{Leiner}, E., {Mathieu}, R.~D., \& {Geller}, A.~M. 2017, \apj, 840, 67

\bibitem[{{Macdonald} \& {Mullan}(2010)}]{macdonald2010}
{Macdonald}, J., \& {Mullan}, D.~J. 2010, \apj, 723, 1599

\bibitem[{{MacDonald} \& {Mullan}(2013)}]{macdonald2013}
{MacDonald}, J., \& {Mullan}, D.~J. 2013, \apj, 765, 126

\bibitem[{{MacDonald} \& {Mullan}(2017)}]{macdonald2017}
---. 2017, \apj, 834, 67

\bibitem[{{Malo} {et~al.}(2014){Malo}, {Doyon}, {Feiden}, {Albert},
  {Lafreni{\`e}re}, {Artigau}, {Gagn{\'e}}, \& {Riedel}}]{malo2014}
{Malo}, L., {Doyon}, R., {Feiden}, G.~A., {et~al.} 2014, \apj, 792, 37

\bibitem[{{Mann} {et~al.}(2015){Mann}, {Feiden}, {Gaidos}, {Boyajian}, \& {von
  Braun}}]{mann2015}
{Mann}, A.~W., {Feiden}, G.~A., {Gaidos}, E., {Boyajian}, T., \& {von Braun},
  K. 2015, \apj, 804, 64

\bibitem[{{Mendoza} {et~al.}(2007){Mendoza}, {Seaton}, {Buerger},
  {Bellor{\'{\i}}n}, {Mel{\'e}ndez}, {Gonz{\'a}lez}, {Rodr{\'{\i}}guez},
  {Delahaye}, {Palacios}, {Pradhan}, \& {Zeippen}}]{mendoza2007}
{Mendoza}, C., {Seaton}, M.~J., {Buerger}, P., {et~al.} 2007, \mnras, 378, 1031

\bibitem[{{Montet} {et~al.}(2017){Montet}, {Tovar}, \&
  {Foreman-Mackey}}]{montet2017}
{Montet}, B.~T., {Tovar}, G., \& {Foreman-Mackey}, D. 2017, \apj, 851, 116

\bibitem[{{Morris} {et~al.}(2018){Morris}, {Curtis}, {Douglas}, {Hawley},
  {Ag{\"u}eros}, {Bobra}, \& {Agol}}]{morris2018}
{Morris}, B.~M., {Curtis}, J.~L., {Douglas}, S.~T., {et~al.} 2018, \aj, 156,
  203

\bibitem[{{Mullan} \& {MacDonald}(2001)}]{mullan2001}
{Mullan}, D.~J., \& {MacDonald}, J. 2001, \apj, 559, 353

\bibitem[{{Pecaut} \& {Mamajek}(2013)}]{pecaut2013}
{Pecaut}, M.~J., \& {Mamajek}, E.~E. 2013, \apjs, 208, 9

\bibitem[{{Pecaut} \& {Mamajek}(2016)}]{pecaut2016}
---. 2016, \mnras, 461, 794

\bibitem[{{Pecaut} {et~al.}(2012){Pecaut}, {Mamajek}, \& {Bubar}}]{pecaut2012}
{Pecaut}, M.~J., {Mamajek}, E.~E., \& {Bubar}, E.~J. 2012, \apj, 746, 154

\bibitem[{{Pinsonneault}(1997)}]{pinsonneault1997}
{Pinsonneault}, M. 1997, \araa, 35, 557

\bibitem[{{Pizzolato} {et~al.}(2003){Pizzolato}, {Maggio}, {Micela},
  {Sciortino}, \& {Ventura}}]{pizzolato2003}
{Pizzolato}, N., {Maggio}, A., {Micela}, G., {Sciortino}, S., \& {Ventura}, P.
  2003, \aap, 397, 147

\bibitem[{{Rackham} {et~al.}(2018){Rackham}, {Apai}, \&
  {Giampapa}}]{rackham2018}
{Rackham}, B.~V., {Apai}, D., \& {Giampapa}, M.~S. 2018, \apj, 853, 122

\bibitem[{{Rebull} {et~al.}(2018){Rebull}, {Stauffer}, {Cody}, {Hillenbrand },
  {David}, \& {Pinsonneault}}]{rebull2018}
{Rebull}, L.~M., {Stauffer}, J.~R., {Cody}, A.~M., {et~al.} 2018, \aj, 155, 196

\bibitem[{{Roettenbacher} {et~al.}(2016){Roettenbacher}, {Monnier}, {Korhonen},
  {Aarnio}, {Baron}, {Che}, {Harmon}, {K{\H{o}}v{\'a}ri}, {Kraus}, \&
  {Schaefer}}]{roettenbacher2016}
{Roettenbacher}, R.~M., {Monnier}, J.~D., {Korhonen}, H., {et~al.} 2016, \nat,
  533, 217

\bibitem[{{Rogers} \& {Nayfonov}(2002)}]{rogers2002}
{Rogers}, F.~J., \& {Nayfonov}, A. 2002, \apj, 576, 1064

\bibitem[{{Rogers} {et~al.}(1996){Rogers}, {Swenson}, \&
  {Iglesias}}]{rogers1996}
{Rogers}, F.~J., {Swenson}, F.~J., \& {Iglesias}, C.~A. 1996, \apj, 456, 902

\bibitem[{{Soderblom} {et~al.}(1993){Soderblom}, {Jones}, {Balachandran},
  {Stauffer}, {Duncan}, {Fedele}, \& {Hudon}}]{soderblom1993a}
{Soderblom}, D.~R., {Jones}, B.~F., {Balachandran}, S., {et~al.} 1993, \aj,
  106, 1059

\bibitem[{{Somers} \& {Pinsonneault}(2014)}]{somers2014}
{Somers}, G., \& {Pinsonneault}, M.~H. 2014, \apj, 790, 72

\bibitem[{{Somers} \& {Pinsonneault}(2015{\natexlab{a}})}]{somers2015a}
---. 2015{\natexlab{a}}, \apj, 807, 174

\bibitem[{{Somers} \& {Pinsonneault}(2015{\natexlab{b}})}]{somers2015b}
---. 2015{\natexlab{b}}, \mnras, 449, 4131

\bibitem[{{Somers} \& {Stassun}(2017)}]{somers2017a}
{Somers}, G., \& {Stassun}, K.~G. 2017, \aj, 153, 101

\bibitem[{{Somers} {et~al.}(2017){Somers}, {Stauffer}, {Rebull}, {Cody}, \&
  {Pinsonneault}}]{somers2017b}
{Somers}, G., {Stauffer}, J., {Rebull}, L., {Cody}, A.~M., \& {Pinsonneault},
  M. 2017, \apj, 850, 134

\bibitem[{{Spruit}(1982)}]{spruit1982}
{Spruit}, H.~C. 1982, \aap, 108, 348

\bibitem[{{Spruit} \& {Weiss}(1986)}]{spruit1986}
{Spruit}, H.~C., \& {Weiss}, A. 1986, \aap, 166, 167

\bibitem[{{Stahler}(1988)}]{stahler1988}
{Stahler}, S.~W. 1988, \apj, 332, 804

\bibitem[{{Stassun} {et~al.}(2014{\natexlab{a}}){Stassun}, {Feiden}, \&
  {Torres}}]{stassun2014a}
{Stassun}, K.~G., {Feiden}, G.~A., \& {Torres}, G. 2014{\natexlab{a}}, \nar,
  60, 1

\bibitem[{{Stassun} {et~al.}(2012){Stassun}, {Kratter}, {Scholz}, \&
  {Dupuy}}]{stassun2012}
{Stassun}, K.~G., {Kratter}, K.~M., {Scholz}, A., \& {Dupuy}, T.~J. 2012, \apj,
  756, 47

\bibitem[{{Stassun} {et~al.}(2014{\natexlab{b}}){Stassun}, {Scholz}, {Dupuy},
  \& {Kratter}}]{stassun2014b}
{Stassun}, K.~G., {Scholz}, A., {Dupuy}, T.~J., \& {Kratter}, K.~M.
  2014{\natexlab{b}}, \apj, 796, 119

\bibitem[{{Stauffer} {et~al.}(2014){Stauffer}, {Cody}, {Baglin}, {Alencar},
  {Rebull}, {Hillenbrand}, {Venuti}, {Turner}, {Carpenter}, {Plavchan},
  {Findeisen}, {Carey}, {Terebey}, {Morales-Calder{\'o}n}, {Bouvier}, {Micela},
  {Flaccomio}, {Song}, {Gutermuth}, {Hartmann}, {Calvet}, {Whitney}, {Barrado},
  {Vrba}, {Covey}, {Herbst}, {Furesz}, {Aigrain}, \& {Favata}}]{stauffer2014}
{Stauffer}, J., {Cody}, A.~M., {Baglin}, A., {et~al.} 2014, \aj, 147, 83

\bibitem[{{Stauffer} {et~al.}(2003){Stauffer}, {Jones}, {Backman}, {Hartmann},
  {Barrado y Navascu{\'e}s}, {Pinsonneault}, {Terndrup}, \&
  {Muench}}]{stauffer2003}
{Stauffer}, J.~R., {Jones}, B.~F., {Backman}, D., {et~al.} 2003, \aj, 126, 833

\bibitem[{{Torres} {et~al.}(2010){Torres}, {Andersen}, \&
  {Gim{\'e}nez}}]{torres2010}
{Torres}, G., {Andersen}, J., \& {Gim{\'e}nez}, A. 2010, \aapr, 18, 67

\bibitem[{{Vorobyov} \& {Basu}(2005)}]{vorobyov2005}
{Vorobyov}, E.~I., \& {Basu}, S. 2005, \apjl, 633, L137

\bibitem[{{Vorobyov} \& {Basu}(2015)}]{vorobyov2015}
---. 2015, \apj, 805, 115

\bibitem[{{Yadav} {et~al.}(2015){Yadav}, {Christensen}, {Morin}, {Gastine},
  {Reiners}, {Poppenhaeger}, \& {Wolk}}]{yadav2015}
{Yadav}, R.~K., {Christensen}, U.~R., {Morin}, J., {et~al.} 2015, \apjl, 813,
  L31

\end{thebibliography}

\end{document}